\UseRawInputEncoding
\providecommand{\bibinfo}[2]{#2}
\documentclass[aip,jcp,
 amsmath,amssymb,
 reprint, %
]{revtex4-1}
\usepackage{graphicx}
\usepackage{dcolumn}
\usepackage{bm}

\usepackage[utf8]{inputenc}
\usepackage[T1]{fontenc}
\usepackage{mathptmx}
\usepackage[normalem]{ulem}
\usepackage{etoolbox}

\usepackage{xcolor}
\usepackage{url}
\usepackage{cleveref}
\usepackage{newunicodechar}
\newunicodechar{⁻}{$^{-}$}
\usepackage{url}
\usepackage{placeins}
\usepackage[english]{babel}
\usepackage{multirow}
\usepackage{newunicodechar}
\newunicodechar{−}{-}
\usepackage{booktabs}
\usepackage[utf8]{inputenc}
\usepackage[T1]{fontenc}
\usepackage{tabularx}
\usepackage{tikz}
\usepackage{array}
\usepackage{makecell}
\usepackage{ragged2e}
\renewcommand{\thetable}{\arabic{table}}
\usepackage{subfigure}

\usepackage{txfonts}
\usepackage{soul} 
\usepackage[version=4]{mhchem}
\makeatletter
\def\@email#1#2{%
 \endgroup
 \patchcmd{\titleblock@produce}
  {\frontmatter@RRAPformat}
  {\frontmatter@RRAPformat{\produce@RRAP{*#1\href{mailto:#2}{#2}}}\frontmatter@RRAPformat}
  {}{}
}%
\makeatother

\begin{document}

\preprint{AIP/123-QED}

\title{Ab Initio Characterization of \ce{C2H4N2} Isomers: Structures, electronic energies, spectroscopic parameters and formation pathways}
\author{Oko Emmanuel Godwin}
 \affiliation{Universidad Autónoma de Chile, Facultad de Ingenieria, Instituto de Ciencias Aplicadas, Nucleo de Astroquimica \& Astrofisica, Av. Pedro de Valdivia 425, Providencia, Santiago, Chile}
\author{Natalia Inostroza}%
 \email{natalia.inostrozapino@gmail.com}
 
\affiliation{Universidad Autónoma de Chile, Facultad de Ingenieria, Instituto de Ciencias Aplicadas, Nucleo de Astroquimica \& Astrofisica, Av. Pedro de Valdivia 425, Providencia, Santiago, Chile
}%

\author{Diego Mardones}
 
\affiliation{Universidad de Chile, Facultad de Ciencias Físicas y Matemáticas, Departamento de Astronomía, Camino el Observatorio 1515, Las Condes, Santiago, Chile}

\author{Luca Bizzocchi}
 
\affiliation{Dipartimento di Chimica “Giacomo Ciamician”, via P. Gobetti 85, Università di Bologna, Bologna, Italy}

\author{Edgar Mendoza}
 
\affiliation{Dept. Ciencias Integradas, Facultad de Ciencias Experimentales, Centro de Estudios Avanzados en F\'{\i}sica, Matem\'atica y Computaci\'on, Unidad Asociada GIFMAN, CSIC-UHU, Universidad de Huelva, Spain}

\author{María Luisa Senent}
 
\affiliation{Departamento de Química y Física Teóricas, Instituto de Estructura de la Materia, IEM-CSIC, 28006 Madrid, Spain; Unidad Asociada GIFMAN, CSIC-UHU, Universidad de Huelva, 21071 Huelva, Spain}

\author{Miguel Carvajal}

\affiliation{Dept. Ciencias Integradas, Facultad de Ciencias Experimentales, Centro de Estudios Avanzados en F\'{\i}sica, Matem\'atica y Computaci\'on, Unidad Asociada GIFMAN, CSIC-UHU, Universidad de Huelva, Spain \\ 
Instituto Universitario Carlos I de Física teórica y computacional, Universidad de Granada, Spain}

\date{\today}

\begin{abstract}

 This work presents a comprehensive theoretical investigation of key isomers of \ce{C2H4N2} using state-of-the-art quantum chemical methods. The objective is to characterize their molecular structures, spectroscopic constants, and electronic energies, and elucidate plausible formation and destruction pathways, providing data critical for astrochemical and atmospheric detection.
 High-accuracy \textit{ab initio} methods were employed, notably CCSD(T)-F12/cc-pVTZ-F12 for optimized geometries. Additional calculations were performed at the CCSD(T)/aug-cc-pVTZ, CCSD(T)/cc-pVTZ, MP2/aug-cc-pVTZ, and CIS levels. Intrinsic reaction coordinate (IRC) calculations were performed at the B3LYP / 6-31G(d, p) level to explore reaction pathways. The ZPE-corrections were determined for all of the isomers considered. Six low-energy C$_2$H$_4$N$_2$ isomers were identified,  all within 1 eV of the global minimum. Among them, methylcyanamide (MCA) exhibits the lowest relative energy ($\sim$0.2 eV) and a significant electric dipole moment of 5.00 D, making it a strong candidate for detection in the gas phase environments. The rotational constants for MCA, computed at the level of CCSD(T)-F12/cc-pVTZ-F12, are $A_e = 34{,}932.44$ MHz, $B_e = 4{,}995.31$ MHz and $C_e = 4{,}520.30$ MHz. The $V_3$ torsional barrier was found to be 631.19 cm$^{-1}$. Centrifugal distortion constants were computed up to sextic order for all isomers. Formation pathways for MCA ---such as CH$_3$N + HCN $\rightarrow$ CH$_3$NHCN--- and related isomers were characterized. The combination of large dipole moments and distinct rotational signatures supports the detectability of methylcyanamide and related C$_2$H$_4$N$_2$ isomers via radioastronomy, IR and MW spectroscopy. Isomerization and reaction pathways involving radical neutral and neutral neutral processes were found to be key to their formation in the gas phase environments. These results offer a robust foundation for future observational and modeling efforts.
\end{abstract}

\maketitle

\section{Introduction} 
\label{intro}

 In the atmosphere, N-bearing molecules such as amines (R$-$NH$_2$) and nitriles (R$-$C$\equiv$N) either in the gas phase or particle-phase find their ways into the atmosphere via sources such as industrial wastes, biomass burning, vehicle exhaust fumes, carbon capture and storage (CCS) technologies, agricultural waste, animal husbandry, etc. \citep{Ali2024}. Measurements indicate that the amount of amines in the Earth's atmosphere has increased over the years enhancing aerosol formation and growth rates, aerosol hygroscopicity, and the activation of cloud condensation nuclei \cite{Kanawade2025} It is now known that these sources cause amine emissions that lead to rich chemistry in the gas phase, which produces intermediates and conversion products converting amines to amides, imines and many other products known to contribute substantially to global climate effects, global mortality rate, diminishing agricultural production, toxicity, etc. \citep{Nielsen2012,Ali2024}. Under this context, \citet{Nielsen2012} showed that reactions between OH radical and alkyl amine occur very quickly, with a rate coefficient > $10^{11} \, \text{cm}^3 \, \text{molecule}^{-1} \, \text{s}^{-1}$. These reactions contribute to a major sink of alkylamines in the gas phase and typically they diminish within 6 hours in the atmosphere.  
 
 Reactive N-bearing species such as methylamidogen radical CH$_3$N$^{\cdot}$H with $^{\cdot}$NO and others bring about the formation of intermediates or products responsible for the cascading effects of N-bearing species and/or the variation in observed abundances of N-bearing species compared to models \citep{Galloway2003-ob,Suzuki2018-yg,Rezende2021-qi}. 
 
 Beyond Earth atmosphere, N-bearing species are important for astrophysical environments as they are very abundant in the Taurus molecular cloud \citep{Chen2022}.  Thus, the study of such molecular species, reaction processes and the spectral characterization of N-bearing compounds plays an important role in  understanding the formation processes of other prebiotic/biological molecules and learn about the physical conditions of the environments where they are formed, such as exoplanet atmosphere and the interstellar medium (ISM) \citep{Carvajal2019,Bogelund2019,Canelo2021,Chuang2022,Carvajal2024}. Hence the accurate investigation of their structures, spectroscopy, and formation pathways are prerequisites for successful research.

   In the interstellar medium, most N-bearing compounds such as hydrogen cyanides (HCN), cyanamide (\ce{NH2CN} and its tautomeric form HNCNH) are believed to be precursors of purines, pyrimidines and nucleobases~\citep{RamalOlmedo2023}. These types of compounds are very reactive due to the presence of the nucleophillic amine and the electrophillic nitrile groups. Most of the isomers of the general formula \ce{C2H4N2} fall into this group (amines, nitriles or both).

   Apart from the well known isomer aminoacetonitrile (AAN, \ce{NH2CH2CN}) which has been studied extensively and observed in the interstellar medium (ISM) \citep{Melosso2020,Manna2022}, not much is known about the isomer methylcyanamide (\ce{CH3NHCN}, MCA or C1 hereafter).  MCA could be a possible intermediate species in the formation of other N or CN-bearing compounds. There is only a spectral measurement published by \citet{bak1980microwave} although no many assignments are reported. From the spectral assignments, only the rotational constants $B$ and $C$ could be  determined with 0.07~MHz accuracy and, therefore, the structure was assumed from other related molecules such as \ce{CH3NHCl}. Without a good description of the structural and spectroscopic parameters, identification is challenging \citep{inostroza2020theoretical}.
   In addition, the vibrational frequencies and spectroscopic constants of 1,2-diiminoethane (\ce{NHCHCHNH)} was recently investigated as promising molecules for astronomical detection \citep{McKissick2025}. These species fall under the general formula \ce{C2H4N2} as the only known investigated species.
   
Theoretical \textit{ab initio} methods become highly pertinent to describe the rotational structure and the vibrational interactions of these molecules as their anharmonic effects are significantly contributing to the observed spectra. According to  \citet{NAIR2006137,INOSTROZA201225,McCarthy2020} these constants such as rotational constants, centrifugal distortion constants, dipole moments, etc. are crucial to achieve accurate structural description and, in extension, enable potential detection of these species in the gas phase environments~\citep{Senent2016}.
Currently, there is dearth of information available for these species in the literature. In particular, MCA as an isomer of the family \ce{C2H4N2} also exhibits large amplitude motions (LAMs) associated with their lowest vibrational mode \citep{bak1980microwave} and, as such, it is important to investigate MCA and the other isomers of this family providing parameters that best describe them. This could be the case of MCA, because it exhibits LAM motions (such as the NH inversion motion and the methyl rotation) which descriptions constitute a challenge~\citep{bak1980microwave}.

 In this contribution, we aimed to carry out high-quality \textit{ab initio} molecular structure calculations for all isomers in their singlet ground electronic state ($^1A_1$) as well as electronic excitation energies for the six most stable isomers.
 We investigated the potential energy surface of \ce{C2H4N2} including several novel isomers and some conformers. All isomers reported in this work are new additions to the literature except for aminoacetonitrile (AAN) and diiminoethane, which are added to the current work for comparison. We computed spectroscopic constants and pathways of formation for some key isomers of low energy. We provide, and compare when available, spectroscopic parameters such as vibrational frequencies, rotational constants, centrifugal distortions constants, dipole moments, and new pathways of formation and destruction of the three low energy isomers of close identity. 
 
This paper is organised as follows: Section \ref{intro} is the introduction of N-bearing species in the atmosphere and the ISM, narrowing down to the \ce{C2H4N2} isomers, along with the aims and motivations for this study. The next section \ref{sec-meth} covers the methods used. Our findings and discussion are presented in section \ref{results} covering structural characterisations, electronic energies, spectroscopic parameters, reaction pathways, the implications of the studied molecules and their formation and destruction pathways in atmospheric and astrophysical contexts, and the last section \ref{concl} is the conclusion.
\section{Computational Methods} \label{sec-meth}

%
  The electronic structure and energy calculations of the molecular species \ce{C2H4N2} were carried out with the explicitly correlated coupled cluster theory with single and double substitutions and a perturbative correction for the connected triple excitations CCSD(T)-F12~\citep{Adler2007-xo,Knizia2009-kh} using Molpro 2015.1 version~\citep{MOLPRO2015}. Throughout this paper, all reported CCSD(T)-F12 energies correspond to the CCSD(T)-F12b.

  The F12 approach accelerates basis-set convergence, providing near-CBS energies without requiring augmented F12 basis sets \cite{Adler2007}, thus explicitly correlated Dunning basis set\cite{Dunning1989} cc-pVTZ-F12 was employed in this work as it provides a more efficient, lower-cost option that still offers good accuracy for many ground-state calculations and designed for CCSD(T)-F12 \cite{Kruse2020}.  Other levels of theory were also considered, taking into account the required accuracy and the available computational resources. All other theoretical calculations, i.e., MP2  \citep{Mller1934} and CCSD(T) \citep{Raghavachari1989},  were performed using both Gaussian 09 and Gaussian 16 suites of programs~\citet{Gaussian2009,g16} as well as Automaton \citep{Yaez2018}, which was used to search for the local minima of isomers (and conformers) along the \ce{C2H4N2} potential energy surface (PES). Vertical excitation calculations were conducted using single excitation configuration interaction (CIS)~\citep{Shu2015} to analise the excitation energies of the first six (6) most stable isomers. 
  
  Vibrational analysis, including anharmonic correction terms,  of the fundamental frequencies of the molecular species \ce{C2H4N2} was performed at CCSD(T)/cc-pVTZ \citep{Dunning89} and MP2/aug-cc-pVTZ levels of theory~\citep{Mller1934} for low-lying isomers. PES of the large amplitude motions of the methylcyanamide (MCA, \ce{CH3NHCN)} was characterised at MP2/aug-cc-PVTZ level of theory. Rotational constants and dipole moments ($\mu$) of \ce{C2H4N2} isomers (and conformers) were calculated at corresponding equilibrium geometries with CCSD(T)-F12/cc-pVTZ-F12,  CCSD(T)/cc-pVTZ and MP2/aug-cc-pVTZ methods. 
  
Reactions leading to the formation and destruction of MCA and other close-lying isomers were investigated within the framework of the Intrinsic Reaction Coordinate (IRC)~\citep{Fukui1970,Fukui1981,Gonzalez1989-ao,Maeda2014} approach using the default settings in Gaussian 09 and Gaussian 16. B3LYP/6-31G(d,p) level of theory was used to compute the IRC pathways due to its good compromise between accuracy and computational cost~\citep{Tirado-Rives2008-xd}. Single-point energy calculations of the transition states were performed with the high-level coupled cluster CCSD(T)  and MP2 levels of theory using a large augmented correlation-consistent polarised valence triple-zeta aug-cc-pVTZ basis set\citep{Mackie2011-vn}. The structures at the minima and transition states were investigated using vibrational analysis obtained by B3LYP using analytical second-order derivative along the path connecting the transition state to intermediate, reactants, and products~\citep{Fukui1981}.

\section{Results and Discussion} \label{results}
\subsection{Electronic energies of \ce{C2H4N2} family}
 
\begin{figure*}[ht]
\setlength{\arrayrulewidth}{0pt}
\renewcommand{\arraystretch}{1.2}
\setlength{\fboxsep}{0pt}
\setlength{\fboxrule}{1.5pt}
\begin{tabular*}{\textwidth}{@{\extracolsep{\fill}} l*6{>{\centering\arraybackslash}m{7em}}@{}}
    & \includegraphics[width=7em]{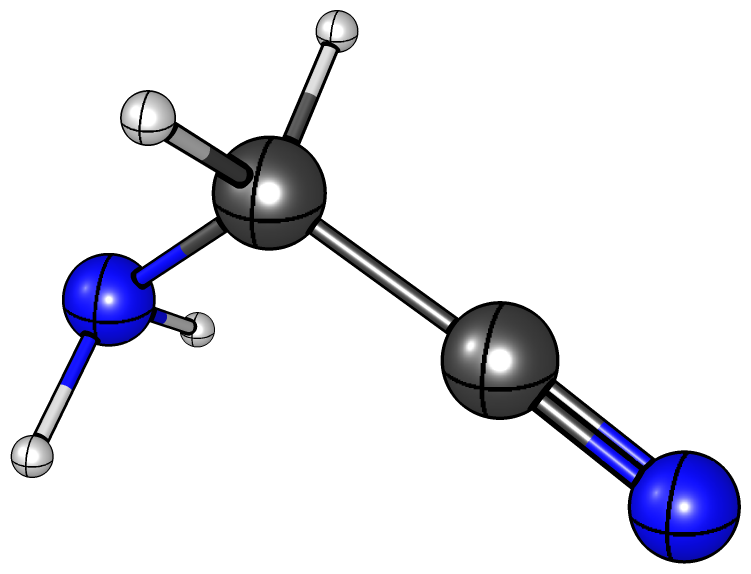} & \includegraphics[width=8em]{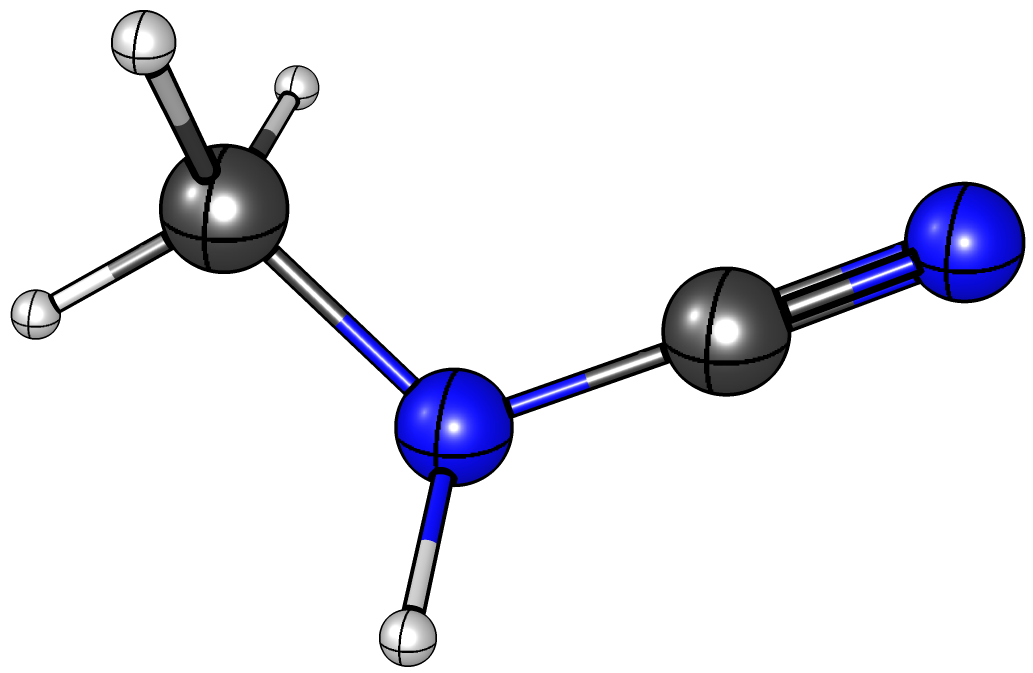} & \includegraphics[width=8em]{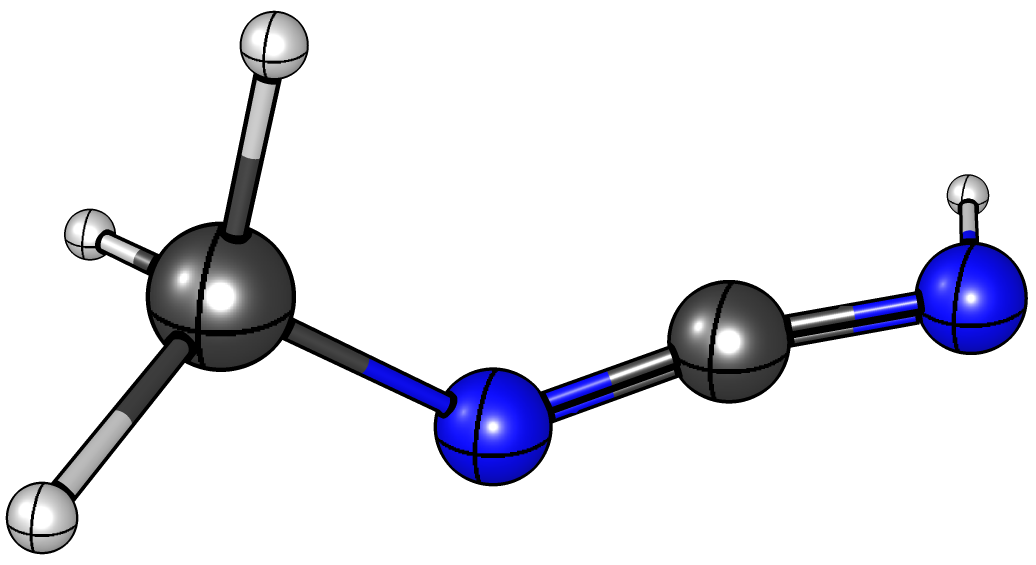} &
    \includegraphics[width=7em]{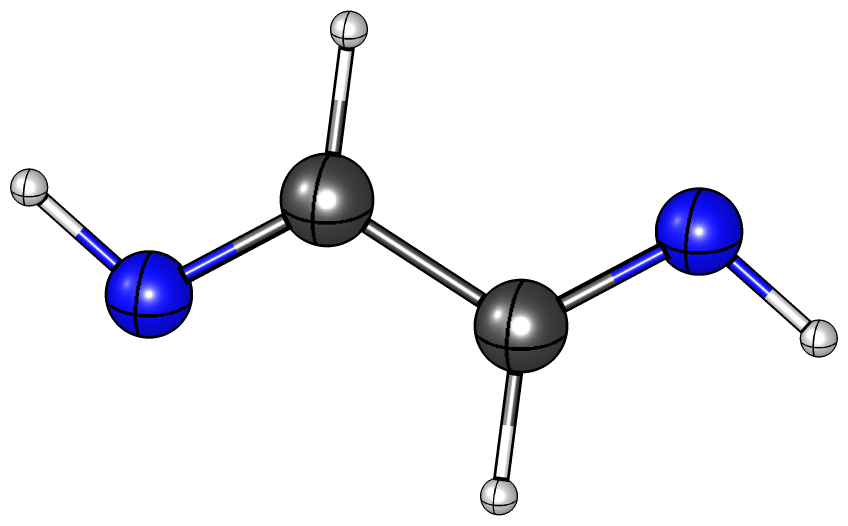} &
    \includegraphics[width=7em]{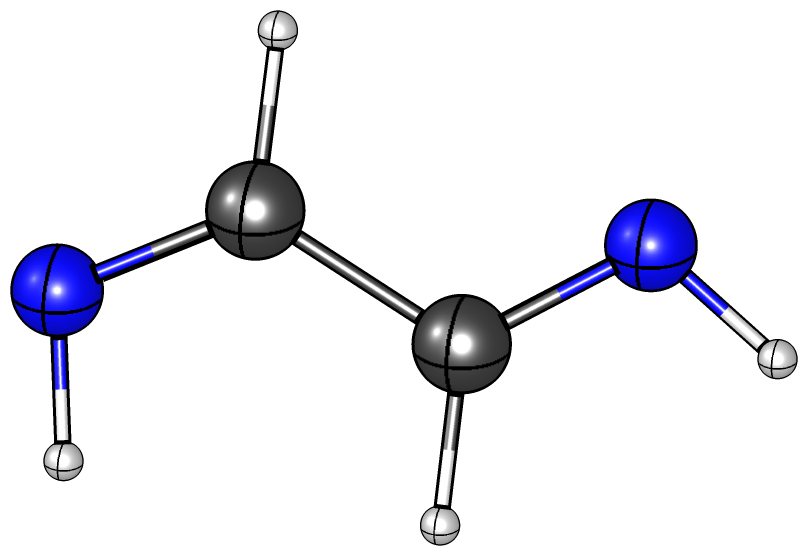} &
    \includegraphics[width=7em]{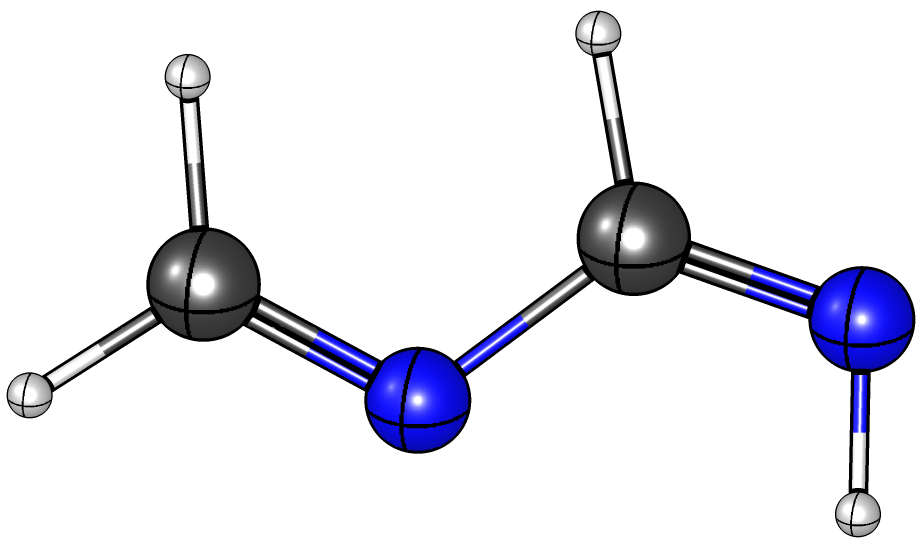}\\
    Label & A1 & C1 & C2 & A2 & A3 & C3 \\ 
    \makecell{Elec. g.s. (P.G.)} & ${}^{1}A'$ (C$_s$) & ${}^{1}A$ (C$_1$) & ${}^{1}A$ (C$_1$) & ${}^{1}A_g$ (C$_{2h}$) & ${}^{1}A'$ (C$_s$) & ${}^{1}A'$ (C$_s$) \\
    $E_{\rm rel}$ ($\text{cm}^{-1}$) & 0.00  & 2177.7  & 3468.2 & 4392.5 & 4920.0 & 6533.1 \\
    ZPVE ($\text{cm}^{-1}$) & 13979.87 & 13796.61 & 13524.24 & 13817.02 & 13728.13 &13644.07  \\
    $E_{\text{rel}}^{+ZPVE}$ ($\text{cm}^{-1}$) & 0.00 & 1994.44 & 3012.57 & 4229.64 & 4668.26 & 6197.30 \\
    $\mu$ (D) &  2.85  & 5.00 & 3.40 & 0.00 &  3.00  & 3.58  \\
    & \includegraphics[width=7.5em]{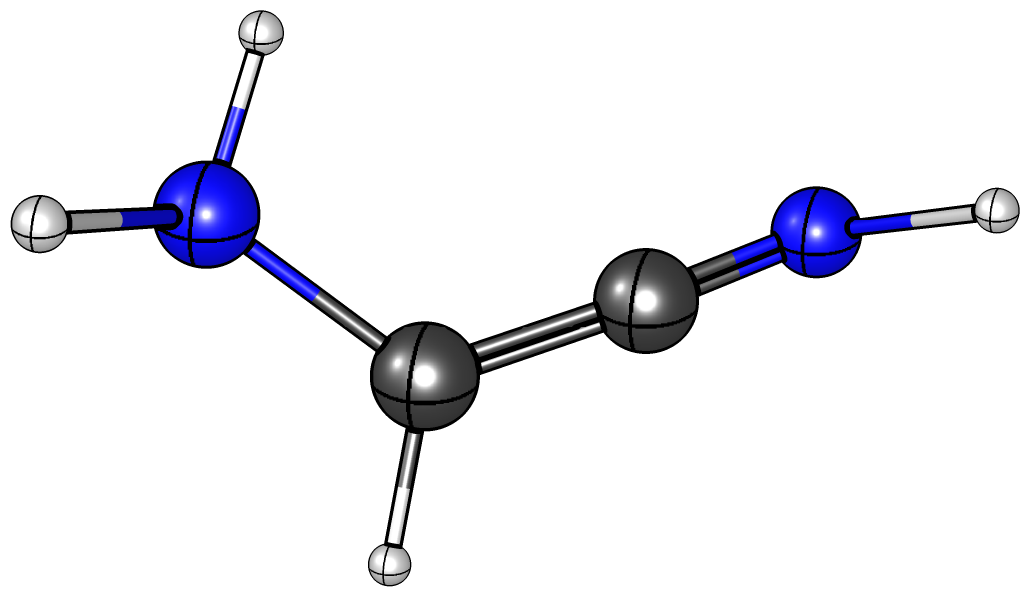} &
    \includegraphics[width=7em]{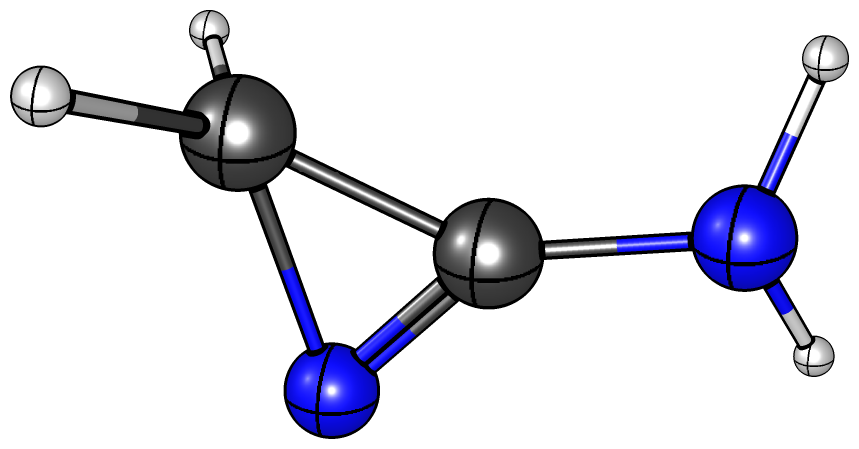} &
    \includegraphics[width=7em]{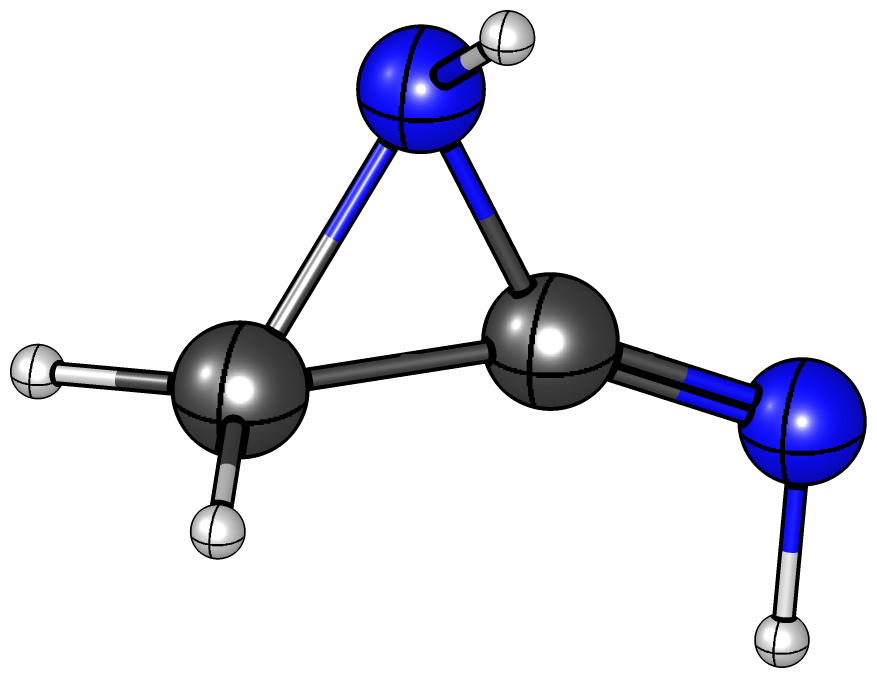} & 
    \includegraphics[width=7em]{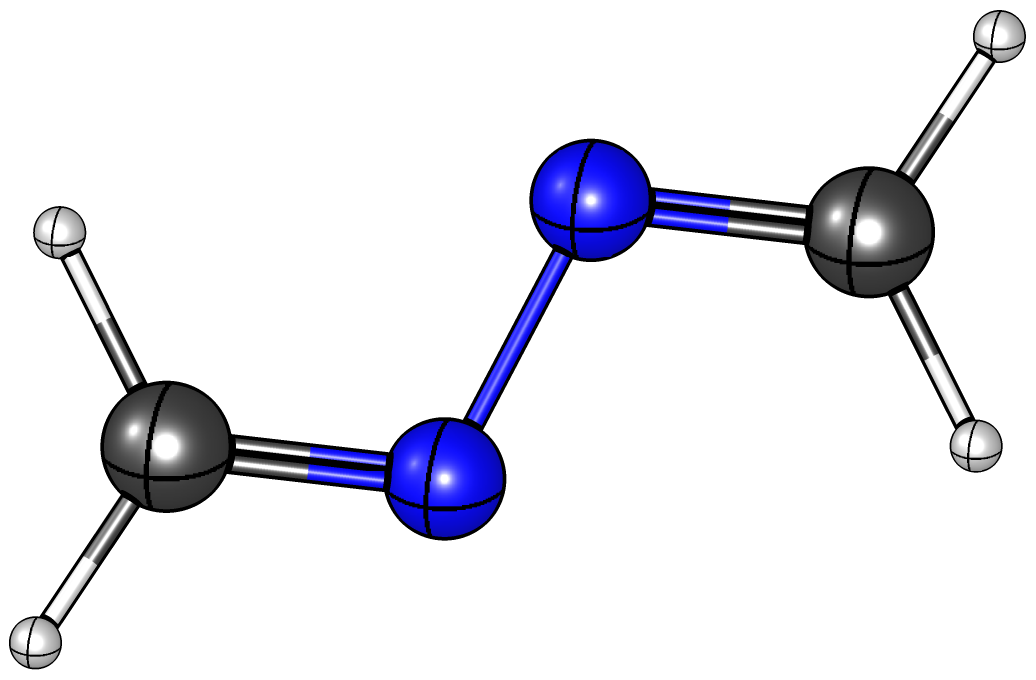} &
    \includegraphics[width=7em]{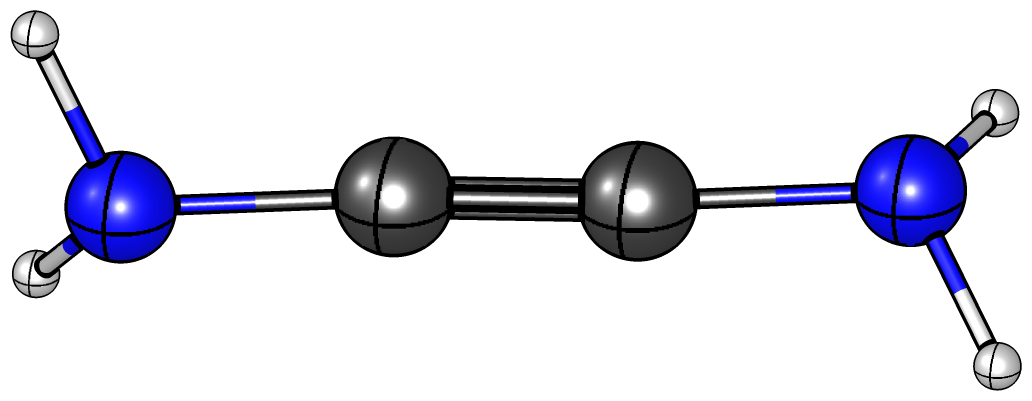} &
    \includegraphics[width=7em]{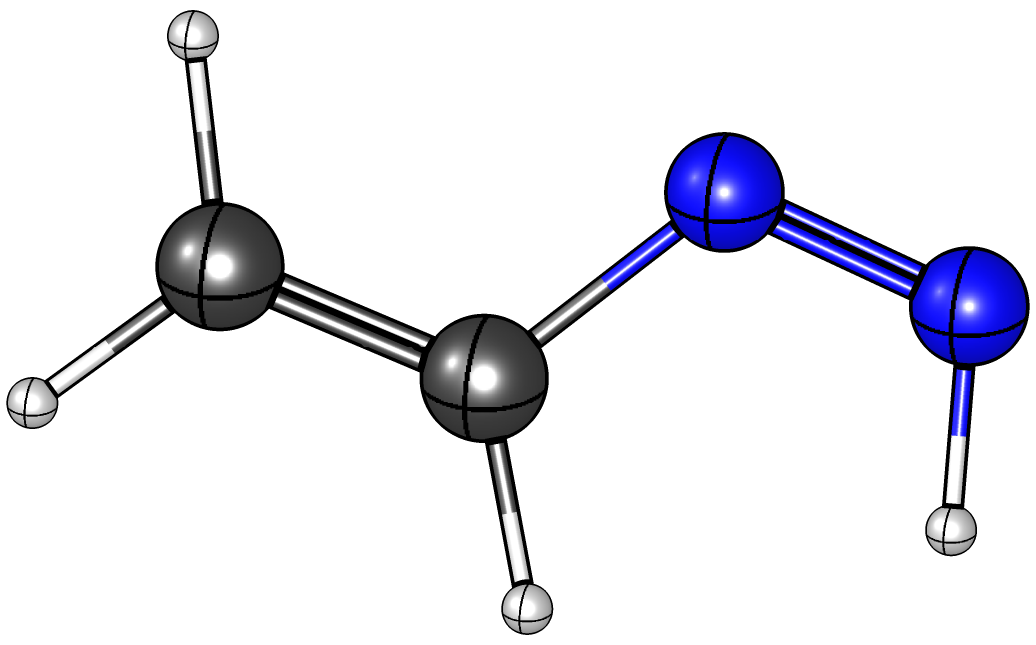}\\
    Label & A4 & C4 & C5 & C6 & A5 & C7 \\ 
    \makecell{Elec. g.s. (P.G.)} & ${}^{1}A$ (C$_1$) & ${}^{1}A$ (C$_1$) & ${}^{1}A$ (C$_1$)
 & ${}^{1}A_g$ (C$_{2h}$) & ${}^{1}A$ (C$_1$)
 & ${}^{1}A'$ (C$_s$) \\
    $E_{\rm rel}$ ($\text{cm}^{-1}$) & 10001.3  & 10323.9 & 12293.4  & 12297.5  & 14025.9 & 14356.7  \\ 
    ZPVE ($\text{cm}^{-1}$) & 13603.47 & 13795.29 & 13842.91 & 13439.30 & 13449.18 & 13432.06 \\
    $E_{\text{rel}}^{+ZPVE}$ ($\text{cm}^{-1}$) & 9624.90 & 10139.32 & 12156.67 & 11756.93 & 13889.17 & 13808.89 \\
    $\mu$ (D) & 2.60 &  3.87 & 3.31  & Nonpolar 
  &  1.70 &  3.22 \\ &
     \includegraphics[width=7em]{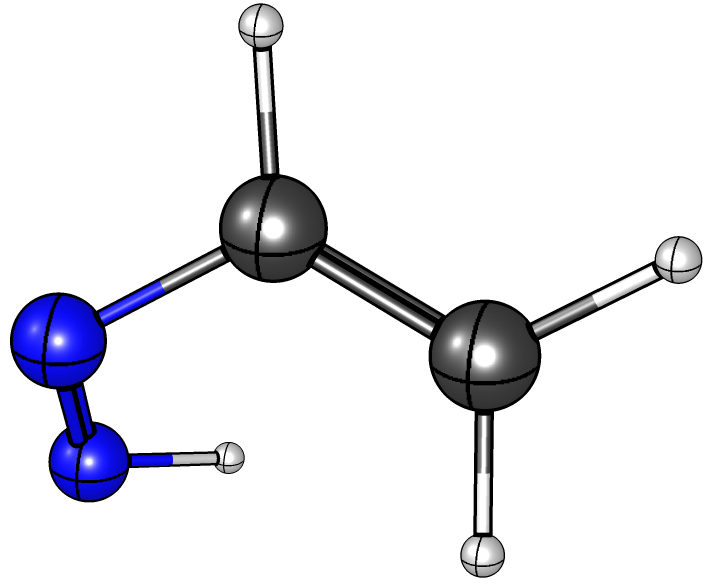} & \includegraphics[width=7em]{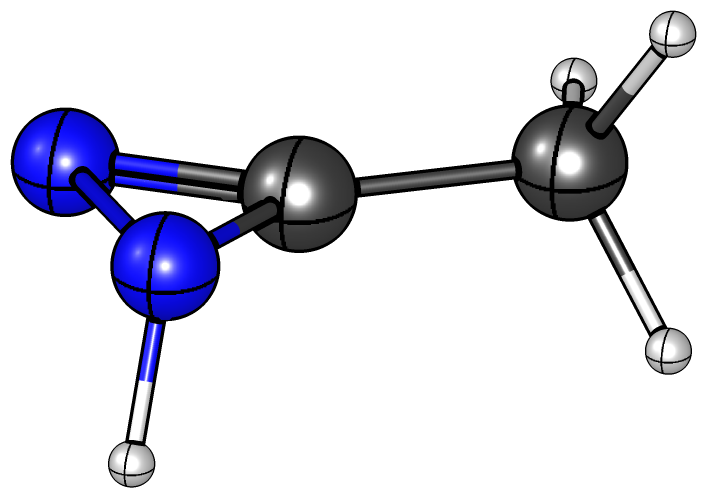} & 
    \includegraphics[width=8em]{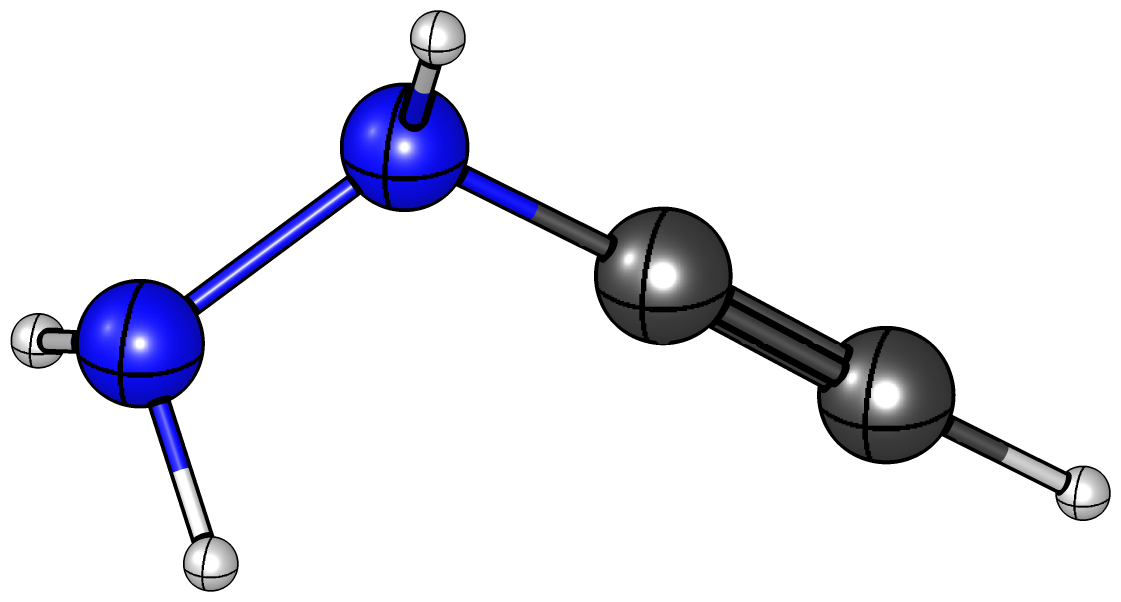} & 
    \includegraphics[width=7em]{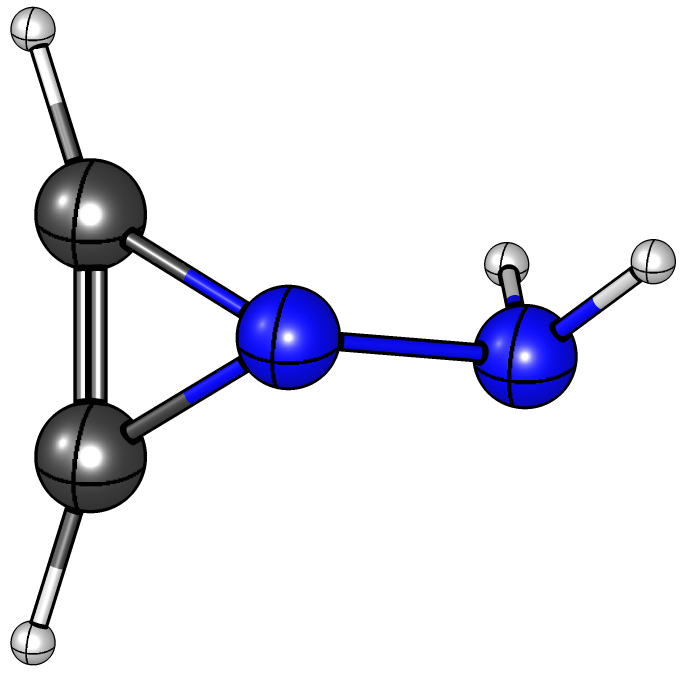} &
    \tikz[baseline={(0,0)}] \node[draw=black, dash pattern=on 3pt off 3pt, thick, inner sep=0pt]{\includegraphics[width=7em]{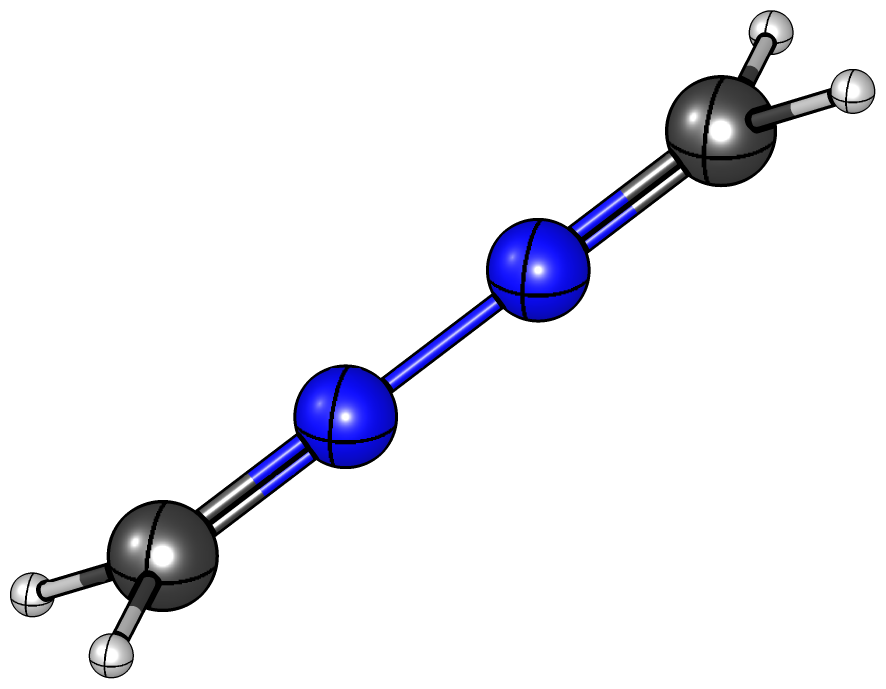}}; &
    \tikz[baseline={(0,0)}] \node[draw=black, dash pattern=on 3pt off 3pt, thick, inner sep=0pt]{\includegraphics[width=7em]{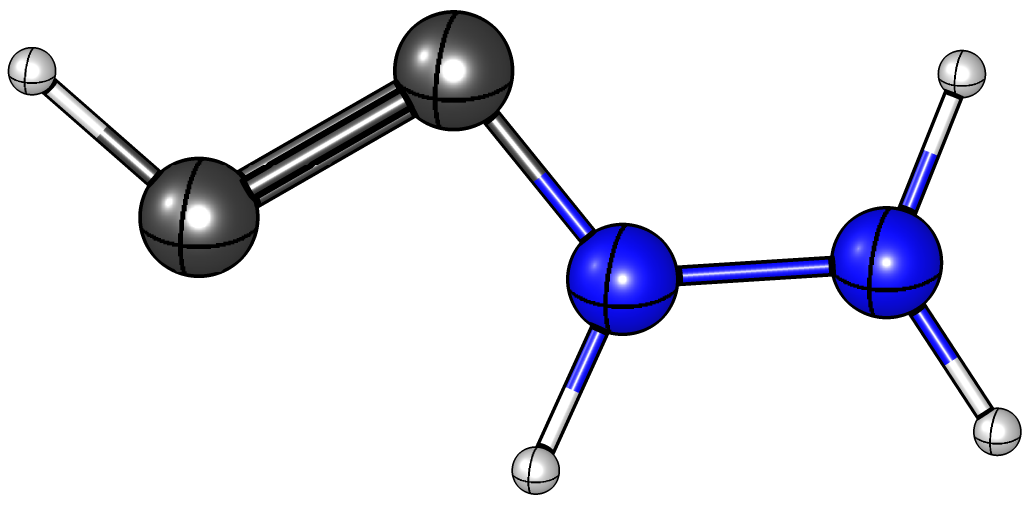}};\\
    Label & C8 & C9 & C10 & C11 & C12 & C13 \\ 
    \makecell{Elec. g.s. (P.G.)} & ${}^{1}A$ (C$_1$) & ${}^{1}A$ (C$_1$) & ${}^{1}A$ (C$_1$) & ${}^{1}A$ (C$_1$) & ${}^{1}A_g$ (D$_{2h}$) & ${}^{1}A'$ (C$_s$) \\
    $E_{\rm rel}$ ($\text{cm}^{-1}$)& 15283.2 & 18389.4  & 21634.6  & 31939.5  &  51780.8  & 54684.4  \\ 
    ZPVE ($\text{cm}^{-1}$) & 13465.86 & 13355.03 & 13609.84 & 12846.72 & - & - \\
    $E_{\text{rel}}^{+ZPVE}$ ($\text{cm}^{-1}$) & 14769.19 & 17764.56 & 21264.57 & 30806.35 & - & - \\
    $\mu$ (D) &  3.13  & 4.01  &  0.92 & 2.38  &  Nonpolar & 3.99
\end{tabular*}
\caption[Isomers of the formula \ce{C2H4N2} and their properties]{\justifying Isomers of the formula \ce{C2H4N2}, ranked by increasing relative energy $E_{\rm rel}$ (in $\text{cm}^{-1}$), referred to the most stable isomer (A1). Their electronic ground states (g.s.), point groups (P.G.), zero-point vibrational energies ZPVE (in $\text{cm}^{-1}$), relative energies including the ZPVE corrections $E_{\text{rel}}^{+ZPVE}$ (in $\text{cm}^{-1}$), and dipole moments $\mu$ (in Debye) are also shown. These calculations were carried out with the level of theory CCSD(T)-F12/cc-pVTZ-F12. The pairs A2 \&{} A3, C6 \&{} C12, C7 \&{} C8, and C10 \&{} C13 are conformers of the same isomer. Transition state structures are highlighted in dotted frames as C12 and C13.}
\label{fig-isomers}
\end{figure*}

In Fig.~\ref{fig-isomers} structures of eighteen isomers of the chemical formula \ce{C2H4N2} are presented with their equilibrium geometries, energies relative to the most stable isomer $E_{\rm rel}$(cm$^{-1}$), zero-point vibrational energies ZPVE(cm$^{-1}$), relative energies corrected with ZPVE, $E_{\rm rel}^{\rm +ZPVE}$(cm$^{-1}$), and their dipole moments $\mu$(D). All these quantities are calculated using the level of theory CCSD(T)-F12/cc-pVTZ-F12, except the zero-point vibrational energy corrections, which is obtained from the CCSD(T)/cc-pVTZ.
In fact, the $E_{\text{rel}}^{+ZPVE}$ presented in Fig.~\ref{fig-isomers} are calculated using  the following approximate approach which provides accurate corrections for the energies~\citep{Villa2011}:

\begin{equation}
    E_{\text{rel}}^{+ZPVE} = \Delta E^{CCSD(T)-F12} + \Delta E_{ZPVE}^{CCSD(T)} ~~,
\end{equation}

\noindent where the energies are computed at the higher-level explicitly correlated coupled-cluster CCSD(T)-F12/cc-pVTZ-F12, and the zero-point vibrational corrections are obtained from the CCSD(T)/cc-pVTZ level. 

Some of the characteristic features of the compounds presented in Fig.~\ref{fig-isomers} include structural and geometric isomerism. All these structures are singlet electronic ground states deduced from the molecular orbitals and the spin multiplicity. The point groups (P.G.) of these isomers are also given. The isomers in Fig.~\ref{fig-isomers} are labeled as An and Cn (n=1, 2, 3, \dots), when they are aminoacetonitrile derivatives and cyanomethylamine derivatives, respectively:
aminoacetonitrile (AAN or A1), methylcyanamide (MCA or C1), methylcarbodiimide (C2), diiminoethane (A2), trans-diiminoethane (A3), methanimidoylformimidamide (C3), aminoethanimine (A4), 2-aminomethylidenaziridine (C4), 2H-azirine (C5), bent-1,2-diazabutadiene (C6), ethynediamine (A5), trans-ethenediazine (C7), cis-ethenediazine (C8),
3-Methyl-3H-diazirine (C9), trans-hydrazinylethyne (C10), hydrazonoethene (C11),  linear-1,2-diazabutadiene(C12), and trans-hydrazinylethyne (C13).

The relative energy  $E_{\rm rel}$ of the molecular species of \ce{C2H4N2} at the  CCSD(T)-F12/cc-pVTZ-F12 level of theory are given in Fig.~\ref{fig-isomers}. They are relevant to predict the potentially observable isomers under different sources. Here it can be noted that the isomer AAN (also named A1), of which the spectrum has been recorded~\citep{Pickett1973} is in fact the only isomer from \ce{C2H4N2} family that has been observed in the ISM~\citep{Manna2022}. The next isomer in energy, MCA or C1, is only at $0.27$~eV from A1 isomer. According to the minimum energy principle, the most stable isomer is usually the most abundant one~\citep{Lattelais2009,Barone2021}. Thus, in line with our calculations, we expect that MCA ( or C1), the second most stable, should also be observed in the ISM. In addition, we also present other isomers, An and Cn with lower 
$E_{\rm rel}$. 
The relative energies ($E_{\rm rel}$) of six lowest-energy isomers sorted based on increasing order: C2 (0.43 eV) < A2 (0.54 eV) < A3 (0.61 eV) < C3 (0.81 eV), placing all four structures within the 0.4–0.8 eV energy range above the global minimum.

From the family of \ce{C2H4N2}, we can distinguish,  among all the molecular structures shown in Fig.~\ref{fig-isomers}, the pairs A2 and A3, C7 and C8 as conformers, while C6 is stable but the matching pair C12 is a transition state, C10 is stable whereas the matching pair C13 is a transition state. 
The energy gaps between the pairs of conformers C6 and C12 and C10 and C13 are large enough and, hence, other structures are found in-between. 

\begin{figure*}
\hspace{-0.8cm}\includegraphics[width=1.10\textwidth]{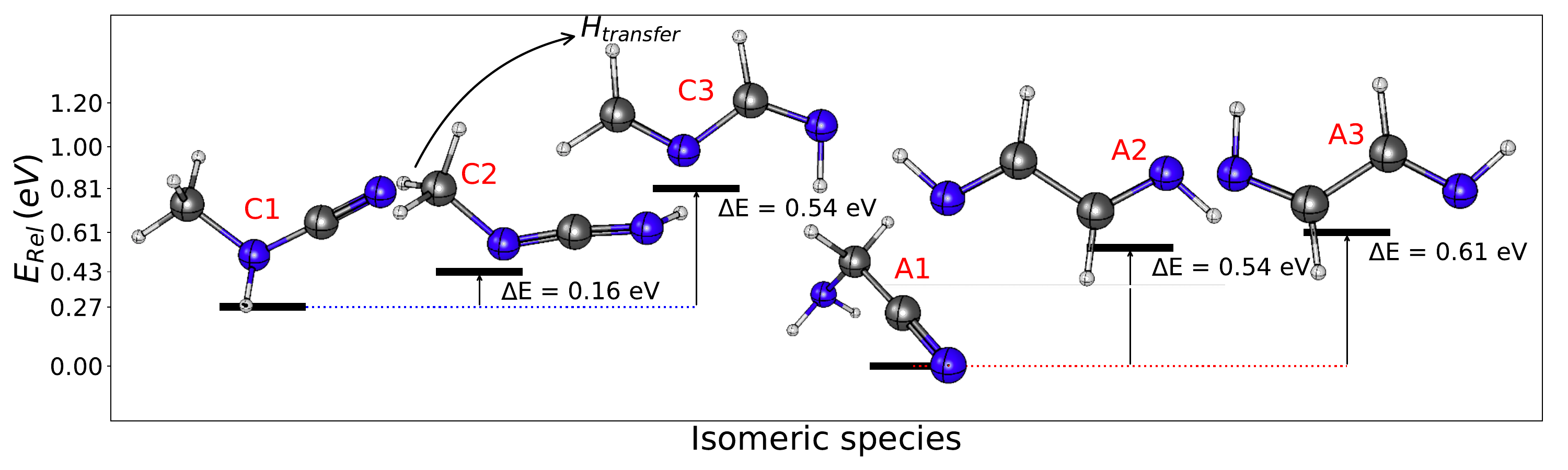}
    \caption{\justifying Proton transfer energy diagram of the two most stable isomers of \ce{C2H4N2} with their relative energies obtained from CCSD(T)-F12/cc-pVTZ-F12. Aminoacetonitrile derivatives are labeled as An while cyanomethylamine derivatives as Cn.}
    \label{trend}
\end{figure*}

In Fig.~\ref{fig-isomers} we showcase the relative electronic energies with ZPVE corrections ($E_{\rm rel}^{\rm +ZPVE}$) for the isomers excluding the Transition State structures (C12 and C13). With the inclusion of ZPVE correction, isomer A1, with absolute energy $E^{\rm +ZPVE}$=-187.726045~$E_h$, is the most stable structure of the family with a difference of $0.25$~eV with respect to C1. The isomers in Fig.~\ref{fig-isomers} are sorted by their relative energy $E_{\rm rel}$ but it can be noted that the order of C5 and C6 and A5 and C7 are interchanged when considering $E_{\rm rel}^{\rm +ZPVE}$. This is due to the close energy of these two isomers and the effect of the ZPVE. 
The absolute energies $E^{\rm +ZPVE}$ of isomers A1, C1, C2, A2, A3 and C3  are within the energy range from -187.73 to -187.69~$E_h$ (around $0.79$~eV) showing their potential proton transfer  interconvertibility (hydrogen shift) from A1 and C1 to their corresponding derivatives at higher energy, respectively. 
The possible proton transfer reactions from the two most stable isomers, AAN and MCA, to the isomers subsequent in energy are exhibited in Fig.~\ref{trend}. In this diagram we show that a relatively low energy, computed using the CCSD(T)-F12/cc-pVTZ-F12 (see Fig.~\ref{fig-isomers}), is needed to isomerise the species A1 and C1 via a proton transfer reaction. The isomerisation energy from A1 to A2 and from A1 to A3 are 52.10~kJ/mol (0.54~eV) and, 58.86~kJ/mol (0.61~eV), respectively. 
For the methylcyanamide derivatives, the formation of isomer C3 proceeds through two sequential steps: C1 → C2 (first proton transfer) and C2 → C3 (second proton transfer). The isomerization energies associated with these transformations are 15.44~kJ/mol (0.16~eV) and  52.10~kJ/mol (0.54~eV), respectively.

 \subsubsection{Comparison among the levels of theory} 
  
\begin{figure}[ht]
    \includegraphics[scale=0.54]{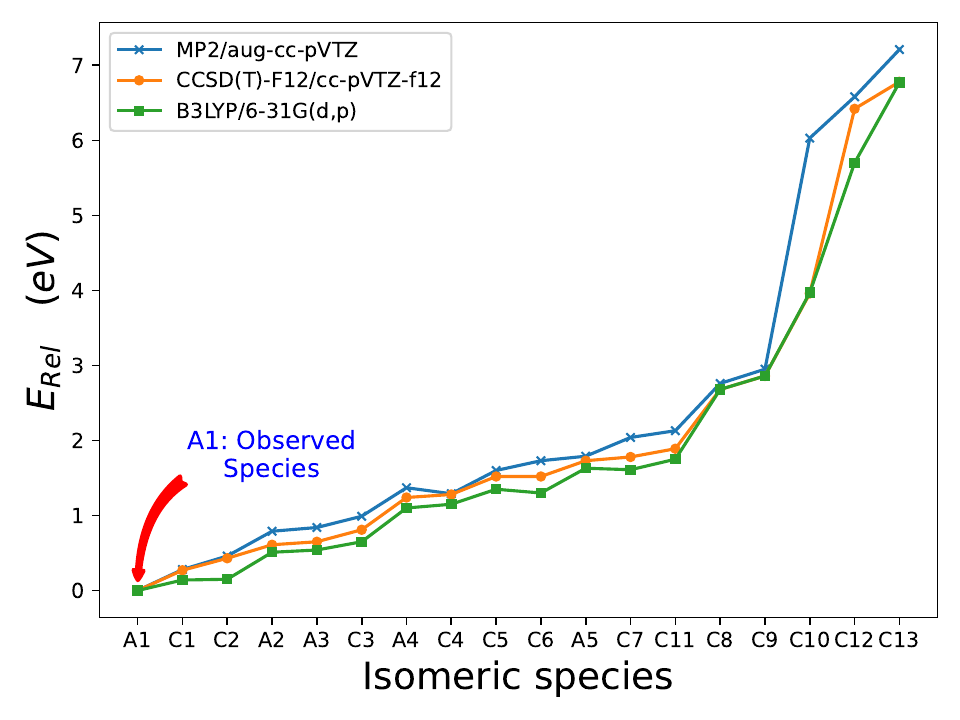}
    \caption{\justifying Comparison of the relative energies $E_{\text{rel}}$ (in eV) calculated with MP2/aug-cc-pVTZ, CCSD(T)-F12/cc-pVTZ-F12, and B3LYP/6-31G(d,p) methods across 18 isomeric species. The relative energies are given with respect to the value of the most stable isomer A1.}
    \label{plot-energy-comp}
\end{figure}

  To evaluate the performance of the three computational methods—MP2/aug-cc-pVTZ, CCSD(T)-F12/cc-pVTZ-F12, and B3LYP/6-31G(d,p)—introduced in Section~\ref{sec-meth}, we compare the relative energies of the eighteen \ce{C2H4N2} isomers under investigation with respect to the lowest-energy structure A1, the only one characterized spectroscopically and observed in the ISM, e.g., in Sgr B2(N)~\citep{Belloche2008,Melosso2020}. As shown in Figure~\ref{plot-energy-comp}, all three levels of theory yield consistent energy hierarchy, with only minor variations among them, highlighting their overall agreement in predicting the relative stabilities of the isomers.
  
  In general, B3LYP provides the lowest energy values and the results of MP2 are closer to the ones obtained by CCSD(T)-F12. In particular, for the species C1, a good agreement is observed between the two levels of theory CCSD(T)-F12/cc-pVTZ-F12 and MP2/aug-cc-pVTZ resulting in a relative energy of 0.27~eV and 0.28~eV, respectively. 

MP2 provides results comparable to the higher level method CCSD(T)-F12,  showing the effectiveness of MP2 level. Furthermore, when MP2 has a divergent behaviour, it can be corrected by including excitation correction terms as  they are implemented in the more robust coupled-cluster approach~\citep{Bochevarov2005}. Particularly, the triple substitutions treated perturbatively in the coupled cluster (CCSD(T)) was considered a gold standard in quantum chemical calculations due to its high accuracy~\citep{Raghavachari1989}. However, the method with the highest accuracy adopted in this study is the explicitly correlated (F12) coupled cluster (CCSD(T)-F12). This method is based on the gold standard and includes germinal basis functions, i.e., a small set of two-particle basis functions that depends explicitly on the inter-electronic distances  and looks almost like the correlation holes.
CCSD(T)-F12 has the advantage of requiring not so large basis sets than for the CCSD(T) as well as it is relatively faster and can be used to predict energy and other chemical properties with a high accuracy~\citep{minkin1997molecular,TEW202183}.

In general, MP2 slightly overestimates the energies with respect to those of CCSD(T)-F12 whereas B3LYP underestimates them. From now on, CCSD(T)-F12 and MP2 will be used, besides estimating the energies, to determine the structure of the isomers especially when high accuracy is needed.

 \subsection{Structural and Spectroscopic parameters}

\subsubsection{Structural parameters}

Table~\ref{tab-geom_xtion_aan_cma_etc} presents the equilibrium structural parameters of selected \ce{C2H4N2} isomers optimized at the CCSD(T)-F12/cc-pVTZ-F12 level of theory. Reported values include bond lengths (in Å) and bond angles (in degrees) for the global minimum structure A1, as well as for the two next-lowest-energy aminoacetonitrile derivatives, A2 and A3.

We have calculated the equilibrium internal coordinates for A1 to validate the {\it ab initio} level of theory used here by comparing them with the experimental and other {\it ab initio} results. It can be noted that the agreement is rather good except for the bond angles $\angle$(C1-C5$\equiv$N8) and $\angle$(H2-C1-H3). On the one hand, the value of $\angle$(C1-C5$\equiv$N8) has a discrepancy of about 5$^{\circ}$ in comparison with the CBS+CV result. Its experimental value was assumed to be fixed at 180$^{\circ}$. On the other hand, angle $\angle$(H2-C1-H3) differs from the experimental value around  5$^{\circ}$. This last difference can be explained because its experimental determination assumed the interbond angle $\angle$(C1-C5$\equiv$N8) as 180$^{\circ}$.
In Table~\ref{tab-geom_xtion_aan_cma_etc-2}, the geometry of  the three lowest energy cyanomethylamine derivatives C1, C2 and C3 are given.
The geometry of the most stable cyanomethylamine derivative C1 is compared with the available experimental interbond angles, which were determined assuming the equilibrium bond lengths r(C7$\equiv$N8) and r(N5-C7) from  another molecule with a similar configuration~\citep{bak1980microwave}. To provide a further comparison, we have also included in Table~\ref{tab-geom_xtion_aan_cma_etc-2} the equilibrium bond lengths and interbond angles, computed at the level of theory B3LYP/aug-cc-pVTZ, of the molecule CH$_3$NH$_2$CN  at the electronic state $^2$A, which is prone to lose one hydrogen of radical NH$_2$~\citep{Sleiman2018}. For this latter case, there is only inconsistency for the $\angle$(N5-C7$\equiv$N8). In general, there is a  good agreement of our calculated parameters with the experimental and theoretical work. In addition, we include the equilibrium structure of the other isomers with higher electronic energy in supplementary information tables: (see Tables~S1 and S2).

\begin{table*}[ht]
\caption{\justifying Equilibrium geometric parameters (bond lengths in $\text{\AA}$ and bond angles in degrees)  computed at the CCSD(T)-F12/cc-pVTZ-F12 level for the first A1, A2, and A3 isomers of C$_2$H$_4$N$_2$.}
\label{tab-geom_xtion_aan_cma_etc}
\begin{tabular}{ccclcccc}
\hline
\multicolumn{3}{c}{A1} & \multicolumn{2}{c}{A2} & \multicolumn{2}{c}{A3} \\ \hline
Description & This work & Exp.$^a$ & CBS+CV$^b$ & Description & This work & Description & This work \\ \hline
R(C5$\equiv$N8) & 1.171 & 1.1594(44) & 1.156 & R(H3N1) & 1.020 & R(H3N1) & 1.024 \\
R(C1C5) & 1.474 & 1.4611(40) & 1.476 & R(N1C2) & 1.281 & R(N1C2) & 1.281 \\
R(C1H2) & 1.089 & 1.0940(48) & 1.088 & R(C2H4) & 1.098 & R(C2H4) & 1.087 \\
R(C1H3) & 1.089 &  &  & R(C2C5) & 1.477 & R(C2C5) & 1.474 \\
R(N4H6) & 1.013 & 1.0138(30) & 1.010 & R(C5H6) & 1.098 & R(C5H6) & 1.095 \\
R(N4-H7) & 1.013 &  &  & R(C5N7) & 1.276 & R(C5N7) & 1.279 \\
R(C1N4) & 1.456 & 1.4760(40) & 1.452 & R(N7H8) & 1.024 & R(N7H8) & 1.020 \\
$\angle$(C1-C5$\equiv$N8) & 177.39 & 180.0 & 182.2 & $\angle$(H3N1C2) & 110.5 & $\angle$(H3N1C2) & 109.82 \\
$\angle$(N4-C1-C5) & 114.79 & 114.54(28) & 114.80 & $\angle$(H4C2C5) & 114.77 & $\angle$(H4C2C5) & 115.77 \\
$\angle$(C1N4H6) & 110.24 & 109.6(5) & 110.5 & $\angle$(H6C5C2) & 114.8 & $\angle$(H6C5C2) & 116.12 \\
$\angle$(C1N4H7) & 110.24 &  &  & $\angle$(H8N7C5) & 110.50 & $\angle$(H8N7C5) & 110.49 \\
$\angle$(H2C1H3) & 107.06 & 102.4(19) &  & $\angle$(C2C5N7) & 119.40 & $\angle$(C2C5N7) & 119.53 \\
$\angle$(H6N4H7) & 106.76 & 107.3(14) &  & $\angle$(H4C2N1) & 125.90 & $\angle$(H4C2N1) & 119.09 \\ \hline
\end{tabular}\flushleft
$^a$ Experimental structural parameters for A   1~\citep{Pickett1973} has been included for validating our results. We have considered the largest uncertainties determined, The angle given without uncertainty is assumed.

$^b$  {\it Ab initio} geometric parameters obtained at the Complete Basis Set (CBS) limit including the core-valence correlation (CV)~\citep{Jiang2023}. The extrapolation was carried out using the levels of theory fc-CCSD(T)/cc-pVnZ level, with n = T and Q, and CCSD(T)/cc-pCVTZ within the fc approximation and correlating all electrons.\\

\end{table*}

\begin{table*}[ht]
\centering
\caption{\justifying Equilibrium geometric parameters (bond lengths in \AA\ and bond angles in degrees)  computed at the CCSD(T)-F12/cc-pVTZ-F12 level for the first C1, C2, and C3 isomers of C$_2$H$_4$N$_2$.}
\label{tab-geom_xtion_aan_cma_etc-2}
\begin{tabular}{ccccccc}
\cline{1-7}
C1 &  &  & \multicolumn{2}{c}{C2} & \multicolumn{2}{c}{C3} \\ \hline
Description & This work & Exp.$^a$ & Description & This work & Description & This work \\ \hline
R(C7$\equiv$N8) & 1.175 &  & R(C2N3) & 1.453 & R(H1C2) & 1.093 \\
R(N5C7) & 1.344 &  & R(N3C4) & 1.218 & R(H3C2) & 1.417 \\
R(C1N5) & 1.467 &  & R(C4N7) & 1.237 & R(N3N4) & 1.279 \\
R(N5H6) & 1.009 &  & R(N7H8) & 1.012 & R(C4H5) & 1.094 \\
R(C1H3) & 1.009 &  & R(C2H1) & 1.089 & R(C4H6) & 1.084 \\
R(C1H2) & 1.086 &  & R(C2H5) & 1.087 & R(N7C2) & 1.271 \\
R(C1H4) & 1.086 &  & R(C2H6) & 1.088 & R(N6H8) & 1.023 \\
$\angle$(N5C7$\equiv$N8) & 176.85 & 179.92 & $\angle$(C2N3C4) & 128.05 & $\angle$(H1C2N3) & 116.95 \\
$\angle$(C1N5C7) & 116.18 & 112.90 & $\angle$(N3C4N7) & 170.39 & $\angle$(H1C2N7) & 119.34 \\
$\angle$(N5C1H4) & 108.21 &  & $\angle$(C4N7H8) & 117.23 & $\angle$(C2N3C4) & 114.84 \\
$\angle$(H4C1H3) & 109.68 &  & $\angle$(N3C2H1) & 112.22 & $\angle$(N3C2N7) & 123.70 \\
$\angle$(H6N5C7) & 112.69 &  & $\angle$(H1C2H6) & 109.33 & $\angle$(H8N7C2) & 108.37 \\
$\angle$(H6N5C1) & 114.54 &  & $\angle$(N3C2H5) & 108.66 & $\angle$(N4C5H7) & 119.09 \\ \hline
\end{tabular}

$^a$ Experimental coordinates determined assuming $\mathrm{r}$(C7$\equiv$N8)=1.160~$\AA$ and $\mathrm{r}$(N5C7)=1.400~$\AA$~\citep{bak1980microwave}. 
\end{table*}

\subsubsection{Harmonic and Anharmonic vibrational Frequencies}

The fundamental frequencies of isomers \ce{C2H4N2} from A1 to C11, ruling out the transition states C12 and C13, have been calculated using two levels of theory, CCSD(T)/cc-pVTZ and MP2/aug-cc-pVTZ. The harmonic frequencies are computed in this work at the level of theory  CCSD(T)/cc-pVTZ. The anharmonic vibrational frequencies are obtained using

\begin{equation}\label{anharm_eqn}
E = \sum_i \omega_i^{\text{CCSD(T)}} \left( \nu_i + \frac{1}{2} \right) + \sum_{i \geq j} x_{ij}^{\text{MP2}} \left( \nu_i + \frac{1}{2} \right) \left( \nu_j + \frac{1}{2} \right)
\end{equation}

\noindent where $\omega_i$ are the harmonic fundamentals, $\nu_i$ and $\nu_j$ are vibrational quanta, and $x_{ij}$ are the anharmonic parameters. The harmonic contribution at CCSD(T)/cc-PVTZ level is corrected with the anharmonic constants derived using MP2/AVTZ force field, as implemented in Refs.~\cite{Toumi2022,Gamez2019}

 In Table~\ref{tab-harmfreq-CCSDT} we present the harmonic CCSD(T)/cc-pVTZ fundamentals and the anharmonic ones calculated with Eq.~(\ref{anharm_eqn}) of aminoacetonitrile (A1), with their vibrational mode descriptions, and of its derivatives A2 and A3. 
Isomer A1 has been included here as a benchmark of the values computed for the new species in this work. The results obtained for A1 are compared with the experimental vibrational frequencies and the theoretical anharmonic frequencies, denoted as CBS+CV/MP2,   obtained at the complete basis set (CBS) limit including the core-valence correlation (CV) and corrected with anharmonic contributions at the ae-MP2/cc-pCVTZ level~\citep{Jiang2023}.  The structure of isomer A1 has a point group ${\cal C}_s$. This group characterises the first 11 vibrational modes with an irreducible representation $A^{\prime}$ and the other 7 modes with a symmetry $A^{\prime\prime}$. From the experimental data,  seven of the lowest  fundamental frequency have been measured with high resolution techniques~\citep{Melosso2020,Jiang2023} and low resolution experimental data of other 10 fundamental frequencies were also reported~\citep{Bak1975}. In general, our anharmonic frequencies calculated with Eq.~(\ref{anharm_eqn}) have a good agreement (within a few cm$^{-1}$) with their corresponding experimental values and CBS+CV/MP2 results.  Nevertheless, large discrepancies with respect to the experimental frequencies  are found between 40-90~cm$^{-1}$ for the modes $\nu_2$, $\nu_3$, $\nu_4$ and $\nu_9$.
Interestingly, the harmonic frequencies  $\omega_3$, $\omega_4$, and $\omega_9$ at CCSD(T)/cc-PVTZ level are closer to the experimental frequencies than the anharmonic ones.

The low-lying A2 conformer belongs to the ${\cal C}_{2h}$ point group. Thus, the A2 conformer exhibits nine Raman-active modes ($7A_g + 2B_g$) and nine IR-active modes ($3A_u + 6B_u$) as arranged in descending order on the tables. The $A_u$ modes correspond to dipole oscillations along the $z$-axis, whereas the $B_u$ modes involve dipole components in the molecular $xy$ plane.  
All irreducible representations of ${\cal C}_{2h}$ are one-dimensional; therefore, all vibrational modes are nondegenerate.  
These symmetry assignments were confirmed by inspection of the computed normal mode displacements. A comparison of the A2 vibrational modes analyzed, ($\nu_{1}$, $\nu_{3}$, $\nu_{4}$, $\nu_{5}$, $\nu_{13}$, $\nu_{15}$, $\nu_{16}$, $\nu_{17}$) show excellent agreement (within 10 cm$^{−1}$) between the anharmonic (this work) and F12–TcCR+DZ levels\cite{McKissick2025}, all other modes have very good consistency. Low-frequency bending or torsional modes (<600 cm$^{−1}$)) exhibit larger relative deviations, typical of floppy or coupled vibrations.The A3 conformer adopts ${\cal C}_s$ symmetry, leading to two irreducible representations: $A^{\prime}$ (symmetric with respect to the mirror plane) and $A^{\prime\prime}$ (antisymmetric).  Accordingly, the first eleven modes are of $A^{\prime}$ symmetry, whereas modes 12–18 correspond to $A^{\prime\prime}$. Both representations are IR and Raman active, with $A^{\prime}$ modes polarized in the molecular plane and $A^{\prime\prime}$ modes polarized perpendicular to it. A comparison of our vibrational frequencies of A2 and A3 with other {\it ab initio} calculations, corresponding to the conformers anti-(E,E) and anti-(E,Z), respectively~\citep{McKissick2025}, shows a good agreement. The calculations given in Ref.~\citep{McKissick2025} resulted from a harmonic frequency analysis computed at CCSD(T)-F12b/cc-pCVTZ-F12 level of theory with relativistic corrections and the anharmonic analysis at CCSD(T)-F12/cc-pVDZ~\citep{McKissick2025}.

Table~\ref{tab-harmfreq-CCSDT-2} presents the harmonic CCSD(T)/cc-pVTZ fundamentals and the anharmonic frequencies calculated with Eq.~(\ref{anharm_eqn})
of methylcyanamide (C1), along with their vibrational mode descriptions, as well as those from its derivatives C2 and C3. The structures of C1 and C2 have a point group, ${\cal C}_1$ and, therefore, all their vibrational modes are totally symmetric. Isomer C3 has a symmetry, ${\cal C}_{\rm s}$ and their first 13 vibrational modes are $A^{\prime}$ and other 5 (from 14th to 18th) vibrational modes  are $A^{\prime\prime}$. For C1 conformer, \citeauthor{bak1980microwave} reported an "approximate" experimental frequency of 105~cm$^{-1}$ for $\nu_{18}$ mode, closer to the theoretical frequency of 95~cm$^{-1}$ in our work. 
The  harmonic ($\omega_i$) and anharmonic ($\nu_i$) fundamental frequencies for the 6 lowest energy isomers A1, A2, A3, C1, C2 and C3 computed at MP2/aug-cc-pVTZ level of theory  are given in Table \ref{freq-MP2}.  Resonances in Tables \ref{tab-harmfreq-CCSDT}, \ref{tab-harmfreq-CCSDT-2}, and \ref{freq-MP2} were determined from the energy separations between anharmonic vibrational states within about $30 \pm 5~\text{cm}^{-1}$ following Ref.~\cite{Kananenka2018}.

In the supplementary information, the harmonic vibrational frequencies at the level of theory CCSD(T)/cc-pVTZ with the anharmonic MP2/aug-cc-pVTZ corrections calculated with Eq.~(\ref{anharm_eqn}), and the harmonic and anharmonic fundamentals at the level of theory MP2/aug-cc-pVTZ are provided for the isomers with higher electronic energy from A4, omitting the transition states C12 and C13. In Tables~S3 
and S4, the  CCSD(T)/cc-pVTZ harmonic frequencies and the anharmonic fundamentals calculated with Eq.~(\ref{anharm_eqn}) are reported. Table~S5 presents the harmonic and anharmonic fundamentals at the level of theory MP2/aug-cc-pVTZ of A4, C4, C5, C6, A5, C7, C8, C9, C10 and C11 isomers.

\begin{table*}[ht]
\caption{\justifying Harmonic fundamental vibrational frequencies at the CCSD(T)/cc-pVTZ level of theory and the anharmonic frequencies calculated with Eq. (\ref{anharm_eqn}) (both in cm$^{-1}$)  for the isomers A1, A2 and A3.}
\label{tab-harmfreq-CCSDT}
{\footnotesize
\begin{tabular}{llllcllccccclllc}
\hline
\multirow{2}{*}{\begin{tabular}[c]{@{}l@{}}Vib.\\ Mode\end{tabular}} & \multicolumn{1}{c}{A1} &  &  & \multicolumn{1}{l}{} &  &  & A2 & \multicolumn{1}{l}{} & \multicolumn{1}{l}{} & \multicolumn{1}{l}{} & \multicolumn{1}{l}{} & A3 &  &  &\\ \cline{2-16} 
 & \begin{tabular}[c]{@{}l@{}}This Work\\ Harm. Fr\end{tabular} & \begin{tabular}[c]{@{}l@{}}This work\\ Anharm.\end{tabular} & Exp. Freq.$^a$ & \multicolumn{1}{l}{\begin{tabular}[c]{@{}l@{}}CBS + \\ CV/MP2$^d$\end{tabular}} & Description & \begin{tabular}[c]{@{}l@{}}Type of\\ Vibration ($\Gamma_{\rm vib}$)\end{tabular} & \begin{tabular}[c]{@{}c@{}}Harm\\ Freq\end{tabular} & \begin{tabular}[c]{@{}c@{}}Anharm\\ Freq\end{tabular} & Exp.$^{e}$ & \begin{tabular}[c]{@{}c@{}}F12-\\ TcCR+DZ$^{f}$\end{tabular} & \multicolumn{1}{l}{$\Gamma_{\rm vib}$} & \multicolumn{1}{c}{\begin{tabular}[c]{@{}c@{}}Harm\\ Freq\end{tabular}} & \begin{tabular}[c]{@{}l@{}}Anharm\\ Freq\end{tabular} & \begin{tabular}[c]{@{}l@{}}F12-\\ TcCR+\\ DZ$^{f}$\end{tabular} & $\Gamma_{\rm vib}$ \\ \cline{1-16} 
$\nu_1$ & 3518 & 3377 & 3367(1) & 3378 & H6-N4-H7 & Sym. stre ($A^{\prime}$) & 3465 & 3301 &  & 3303 & $A_g$ & 3417 & 3252 & 3303 & $A^{\prime}$\\
$\nu_2$ & 3092 & 2993 & 2950(1) & 2952 & H2-C1-H3 & Sym. stre ($A^{\prime}$) & 3101 & 2948 &  & 2925 & $A_g$ & 3339 & 3169 & 3239 & $A^{\prime}$\\
$\nu_3$ & 2228 & 2185 & 2236(1) & 2254 & C5$\equiv$N8 & Stretching ($A^{\prime}$) & 1700 & 1652 &  & 1652 & $A_g$ & 3157 & 3023 & 2992 & $A^{\prime}$\\
$\nu_4$ & 1673 & 1549 & 1642(1) & 1617 & H6-N4-H7 & Bending ($A^{\prime}$) & 1428 & 1392 &  & 1386 & $A_g$ & 3014 & 2857 & 2881 & $A^{\prime}$\\
$\nu_5$ & 1478 & 1451 & 1444(1) & 1446 & H2-C1-H3 & Bending ($A^{\prime}$) & 1276 & 1249 &  & 1248 & $A_g$ & 1684 & 1638 & 1645 & $A^{\prime}$\\
$\nu_6$ & 1369 & 1334 & 1348(1) & 1338 & H2-C1-H3 & Wagging ($A^{\prime}$) & 1021 & 997 &  & 975 & $A_g$ & 1625 & 1601 & 1616 & $A^{\prime}$\\
$\nu_7$ & 1126 & 1086 & 1077(1) & 1086 & C1-N4 & Stretching ($A^{\prime}$) & 586 & 578 &  & 565 & $A_g$ & 1406 & 1371 & 1381 & $A^{\prime}$\\
$\nu_8$ & 946 & 910 & 901.4(1)$^b$ & 898 & C1-C5 & Stretching ($A^{\prime}$) & 1128 & 1099 &  & 1083 & $A_u$ & 1379 & 1347 & 1361 & $A^{\prime}$\\
$\nu_9$ & 846 & 719 & 790.95908(7)$^b$ & 811 & H6-N4-H7 & Wagging ($A^{\prime}$) & 651 & 634 & 646 & 603 & $A_u$ & 1278 & 1249 & 1259 & $A^{\prime}$\\
$\nu_{10}$ & 571 & 566 & 556.56467(2)$^b$ & 559 & C1-C5-N8 & Bending ($A^{\prime}$) & 169  & 165 &  & 78 & $A_u$ & 1171 & 1146 & 1159 & $A^{\prime}$\\
$\nu_{11}$ & 235 & 235 & 210.575841(5)$^c$ & 211 & N4-C1-C5 & Bending ($A^{\prime}$) & 1103 & 1081 &  & 1056 & $B_g$ & 1029 & 1004 & 996 & $A^{\prime}$\\
$\nu_{12}$ & 3604 & 3437 & 3431(1) & 3448 & H6-N4-H7 & Asym. stre ($A^{\prime \prime})$ & 909  & 891 &  & 878 & $B_g$ & 573 & 566 & 558 & $A^{\prime}$\\
$\nu_{13}$ & 3136 & 2994 & 2975(1) & 2991 & H2-C1-H3 & Asym. Stre ($A^{\prime \prime}$) & 3465 & 3301 &  & 3301 & $B_u$ & 350 & 353 & 348 & $A^{\prime}$\\
$\nu_{14}$ & 1395 & 1355 & 1331(1) & 1360 & H$_2$C1-H$_2$N4 & Twisting ($A^{\prime \prime}$) & 3101 & 2985 & 3114 & 2928 & $B_u$ & 1151 & 1124 & 1107 & $A^{\prime\prime}$\\
$\nu_{15}$ & 1200 & 1174 & 1160(1)$^b$ & 1173 & H2-C1-H3 & Twisting ($A^{\prime \prime}$) & 1640 & 1612 & 1604 & 1613 & $B_u$ & 1112 & 1092 & 1074 & $A^{\prime\prime}$\\
$\nu_{16}$ & 890 & 881 & ---- & 884 & H2-C1-H3 & Rocking ($A^{\prime \prime}$) & 1377 & 1350 &  & 1345 & $B_u$ & 918 & 905 & 895 & $A^{\prime\prime}$\\
$\nu_{17}$ & 392 & 385 & 368.104656(3)$^c$ & 370 & C1-C5-N8 & Bending ($A^{\prime \prime}$) & 1176 & 1152 & 1155 & 1149 & $B_u$ & 668 & 652 & 628 & $A^{\prime\prime}$\\
$\nu_{18}$ & 261 & 254 & 244.891525(3)$^c$ & 248 & H6-N4-H7 & Torsion ($A^{\prime \prime}$) & 337 & 339 &  & 324 & $B_u$ & 186 & 186 & 148 & $A^{\prime\prime}$\\ \hline
\end{tabular}
}

\flushleft
$^a$ Low resolution data of the experimental fundamental frequencies~\citep{Bak1975}. When available, high resolution experimental frequencies are given instead.\\ $^b$ High resolution experimental fundamentals measured by ~\citet{Jiang2023}.\\ $^c$  High resolution experimental fundamentals obtained by~\citet{Melosso2020}.\\    $^d$ {\it Ab initio} anharmonic frequencies, denoted as CBS + CV/MP2, obtained at the Complete Basis Set (CBS) limit including the core-valence correlation (CV) and corrected with anharmonic contributions at the ae-MP2/
cc-pCVTZ level~\citep{Jiang2023}. The CBS extrapolation was carried out using the levels of theory fc-CCSD(T)/cc-pVnZ level, with n = T and Q, and CCSD(T)/cc-pCVTZ within the fc approximation and correlating all electrons. \\
$^e$ Experimental Ar-matrix results from Ref. \cite{Eckhardt2022} \\
$^f$  Calculations from Ref.18, named as F12-TcCR+DZ, resulted from a harmonic contribution computed at the level of theory CCSD(T)-F12b/cc-pCVTZ-F12 with relativistic corrections and the anharmonic one at CCSD(T)-F12/cc-pVDZ. \\
Only interactions between the anharmonic vibrational bands whose energy separation is within about 
$30 \pm 5~\text{cm}^{-1}$ were treated as significant.
; A1: $\nu_{15} \leftrightarrow \nu_{8} + \nu_{18}$,  
A2: $\nu_{14} \leftrightarrow \nu_{16} + \nu_{3}$, $\nu_{15} \leftrightarrow \nu_{18} + \nu_{5}$, $\nu_{16} \leftrightarrow \nu_{18} + \nu_{6}$. 
A3: $\nu_{6} \leftrightarrow \nu_{9} + \nu_{13}$.  

\end{table*}

\begin{table*}[ht]
\centering
\caption{\justifying Harmonic vibrational frequencies at the CCSD(T)/cc-pVTZ level of theory and the anharmonic frequencies computed with Eq.~(\ref{anharm_eqn}) (both in cm$^{-1}$) for the isomers C1, C2 and C3.}
\label{tab-harmfreq-CCSDT-2}
\begin{tabular}{lllllllll}
\hline
\multirow{2}{*}{\begin{tabular}[c]{@{}l@{}}Vib.\\ Mode\end{tabular}} & MCA(C1) &  &  &  & C2 &  & C3 &  \\ \cline{2-9} 
 & \multicolumn{1}{c}{\begin{tabular}[c]{@{}c@{}}Harm.\\ Freq\end{tabular}} & \multicolumn{1}{c}{\begin{tabular}[c]{@{}c@{}}Anharm\\ Freq\end{tabular}} & Description & \begin{tabular}[c]{@{}l@{}}Type of \\ vibration\end{tabular} & \multicolumn{1}{c}{\begin{tabular}[c]{@{}c@{}}Harm\\ Freq\end{tabular}} & \begin{tabular}[c]{@{}l@{}}Aharm\\ Freq\end{tabular} & \multicolumn{1}{c}{\begin{tabular}[c]{@{}c@{}}Harm\\ Freq\end{tabular}} & \begin{tabular}[c]{@{}l@{}}Aharm\\ Freq\end{tabular} \\ \midrule
$\nu_1$ & 3594 & 3436 & N5-H6 & Stretching & 3575 & 3412 & 3435 & 3269 \\
$\nu_2$ & 3176 & 3038 & H2-C1-H4 & Asym. str. & 3129 & 2991 & 3183 & 3047 \\
$\nu_3$ & 3144 & 3008 & CH$_3$ & Asym. str. & 3120 & 2981 & 3070 & 2910 \\
$\nu_4$ & 3060 & 2937 & CH$_3$ & Sym. str. & 3040 & 2981 & 3036 & 2890 \\
$\nu_5$ & 2245 & 2210 & N5-C7-N8 & Stretching & 2202 & 2149 & 1684 & 1646 \\
$\nu_6$ & 1526 & 1480 & CH$_3$ & Asym. bending & 1516 & 1506 & 1667 & 1624 \\
$\nu_7$ & 1501 & 1471 & CH$_3$ & Asym. bending & 1501 & 1472 & 1485 & 1451 \\
$\nu_8$ & 1469 & 1452 & N5-H6 & \begin{tabular}[c]{@{}l@{}}Twisting/\\ bending\end{tabular} & 1454 & 1436 & 1413 & 1380 \\
$\nu_9$ & 1457 & 1416 & CH$_3$ & Sym. bending & 1367 & 1344 & 1281 & 1250 \\
$\nu_{10}$ & 1196 & 1177 & CH$_3$ & Wagging & 1144 & 1116 & 1213 & 1188 \\
$\nu_{11}$ & 1165 & 1139 & C7-N5 & Stretching & 1143 & 1119 & 985 & 957 \\
$\nu_{12}$ & 1146 & 1132 & CH$_3$ & Rocking & 981 & 908 & 605 & 594 \\
$\nu_{13}$ & 936 & 926 & H4-C1-N5 & Bending & 905 & 880 & 364 & 362 \\
$\nu_{14}$ & 656 & 602 & N5-C7-N8 & Bending & 644 & 636 & 1107 & 1083 \\
$\nu_{15}$ & 544 & 516 & N5-C7-N8 & Bending & 586 & 568 & 1070 & 1054 \\
$\nu_{16}$ & 443 & 420 & N5-H6 & Inversion & 473 & 459 & 920 & 903 \\
$\nu_{17}$ & 227 & 224 & C1-N5-C7 & Bending & 197 & 193 & 647 & 626 \\
$\nu_{18}$ & 110 & 95 & CH$_3$ & Torsion & 74 & 61 & 123 & 118 \\ \hline
\end{tabular}
\flushleft
Only interactions between the anharmonic vibrational bands whose energy separation is within about 
$30 \pm 5~\text{cm}^{-1}$ were treated as significant:
C1: $\nu_{4} \leftrightarrow 2\nu_{6}$, $\nu_{8} \leftrightarrow \nu_{13}+\nu_{15}$, 
 $\nu_{12} \leftrightarrow \nu_{13}+\nu_{17}$. 
C2: $\nu_{4} \leftrightarrow 2\nu_{6}$, 
$\nu_{7} \leftrightarrow \nu_{8}+\nu_{18}$, $\nu_{12} \leftrightarrow 2\nu_{16}$, $\nu_{2} \leftrightarrow \nu_{3}$, $\nu_{10} \leftrightarrow \nu_{11}$.
C3: $\nu_{14} \leftrightarrow \nu_{11}+\nu_{18}$. 
 
\end{table*}

\begin{table}[ht]
\caption{\justifying Harmonic ($\omega$) and anharmonic ($\nu$) fundamental vibrational  frequencies (in $\text{cm}^{-1}$) for selected isomers of \ce{C2H4N2} at MP2/aug-cc-pVTZ level of theory$^a$}.
\label{freq-MP2}
\begin{tabular}{crcrrrr}
\hline
\begin{tabular}[c]{@{}c@{}}Vibrat. \\ modes\end{tabular} & \multicolumn{1}{c}{A1} & A2 & \multicolumn{1}{c}{A3} & \multicolumn{1}{c}{C1} & \multicolumn{1}{c}{C2} & \multicolumn{1}{c}{C3} \\ \hline
$\omega_1$ & 3532 & 3485 & 3484 & 3606 & 3602 & 3460 \\
$\omega_2$ & 3108 & 3116 & 3429 & 3199 & 3175 & 3220 \\
$\omega_3$ & 2180 & 1684 & 3168 & 3166 & 3166 & 3091 \\
$\omega_4$ & 1666 & 1409 & 3070 & 3075 & 3074 & 3064\\
$\omega_5$ & 1490 & 1277 & 1673 & 2229 & 2219 & 1674 \\
$\omega_6$ & 1369 & 1015 & 1624 & 1539 & 1528 & 1659 \\
$\omega_7$ & 1123 & 585 & 1405 & 1516 & 1516 & 1486 \\
$\omega_8$ & 941 & 1129 & 1381 & 1470 & 1463 & 1403 \\
$\omega_9$ & 842 & 658 & 1282 & 1463 & 1386 & 1268 \\
$\omega_{10}$ & 562 & 165 & 1177 & 1200 & 1153 & 1201 \\
$\omega_{11}$ & 209 & 1102 & 1023 & 1165 & 1146 & 989 \\
$\omega_{12}$ & 3625 & 910 & 573 & 1152 & 937 & 603 \\
$\omega_{13}$ & 3159 & 3485 & 334 & 938 & 898 & 364 \\
$\omega_{14}$ & 1395 & 3117 & 1145 & 657 & 641 & 1103 \\
$\omega_{15}$ & 1203 & 1628 & 1098 & 532 & 564 & 1074 \\
$\omega_{16}$ & 894 & 1364 & 919 & 436 & 466 & 920 \\
$\omega_{17}$ & 374 & 1169 & 655 & 214 & 188 & 644 \\
$\omega_{18}$ & 255 & 338 & 151 & 167 & 64 & 115 \\ \hline
$\nu_1$ & 3391 & 3320 & 3320 & 3448 & 3440 & 3294 \\
$\nu_2$ & 3010 & 2962 & 3260 & 3062 & 3037 & 3084 \\
$\nu_3$ & 2138 & 1636 & 3036 & 3031 & 3027 & 2931 \\
$\nu_4$ & 1542 & 1372 & 2913 & 2951 & 3015 & 2918 \\
$\nu_5$ & 1463 & 1250 & 1626 & 2195 & 2166 & 1635 \\
$\nu_6$ & 1333 & 991 & 1582 & 1493 & 1519 & 1617 \\
$\nu_7$ & 1083 & 577 & 1371 & 1486 & 1486 & 1452 \\
$\nu_8$ & 905 & 1101  & 1349 & 1453 & 1445 & 1370 \\
$\nu_9$ & 715 & 641  & 1253 & 1421 & 1362 & 1237 \\
$\nu_{10}$ & 557 & 161 & 1152 & 1181 & 1125 & 1176 \\
$\nu_{11}$ & 209 & 1080 & 998 & 1138 & 1121 & 962 \\
$\nu_{12}$ & 3458 & 902 & 566 & 1138 & 864 & 592 \\
$\nu_{13}$ & 3018 & 3320 & 337 & 929 & 873 & 362 \\
$\nu_{14}$ & 1356 & 3001 & 1118 & 603 & 633 & 1078 \\
$\nu_{15}$ & 1178 & 1600 & 1078 & 504 & 546 & 1059 \\
$\nu_{16}$ & 885 & 1337 & 906 & 413 & 451 & 903 \\
$\nu_{17}$ & 367 & 1145 & 639 & 212 & 185 & 623 \\
$\nu_{18}$ & 248 & 340 & 149 & 153 & 50 & 109 \\ \hline
\end{tabular}

\flushleft
 Only interactions between the anharmonic vibrational bands whose energy separation is within about 
$30 \pm 5~\text{cm}^{-1}$ were treated as significant: A1
$\nu_{7} \leftrightarrow 2\nu_{10}$,  
$\nu_{15} \leftrightarrow \nu_{8} + \nu_{18}$. 
\\ A2: $\nu_{16} \leftrightarrow \nu_{6} + \nu_{18}$, $\nu_{15} \leftrightarrow \nu_{5} + \nu_{18}$, $\nu_{14} \leftrightarrow \nu_{3} + \nu_{16}$.
\\ A3: $\nu_{6} \leftrightarrow \nu_{9} + \nu_{13}$.
\\ C1: $\nu_{4} \leftrightarrow 2\nu_{6}$, $\nu_{8} \leftrightarrow \nu_{13}+\nu_{15}$,  
$\nu_{12} \leftrightarrow \nu_{13}+\nu_{17}$.
\\
C2: 
$\nu_{7} \leftrightarrow \nu_{8}+\nu_{18}$, $\nu_{4} \leftrightarrow 2\nu_{6}$,  $\nu_{2} \leftrightarrow \nu_{3}$, $\nu_{10} \leftrightarrow \nu_{11}$.
\\
C3: $\nu_{14} \leftrightarrow \nu_{11}+\nu_{18}$,

\end{table}

\subsubsection{Spectroscopic parameters and dipole moments}

The spectroscopic constants of the family of isomers \ce{C2H4N2} are also made available. In this section, we pay special attention to the spectroscopic constants of the energy lowest aminoacetonitrile derivatives A1, A2 and A3 and cyanomethylamine derivatives C1, C2 and C3 as well as providing their dipole moments. 
The \ce{C2N2H4}  is a polar molecule and exhibits closely spaced conformers, diffuse functions \cite{Dunning89} may influence geometries and relative energies. Due to this, geometry optimizations and harmonic frequencies were performed at the MP2/aug-cc-pVTZ level to  include diffuse functions for structural description and ZPE corrections. Electronic reference energies were computed at the CCSD(T)-F12/cc-pVTZ-F12 level; the F12 approach accelerates basis-set convergence, providing near-CBS energies without requiring augmented F12 basis sets, which are less standardized and can introduce numerical issues\cite{Adler2007}. Selected CCSD(T)/cc-pVTZ single-points were employed to validate the protocol.
A first comparison of the equilibrium rotational constants and the total dipole moments of the six lowest energy isomers is showed in Table~\ref{Dipole_rot_comparison} using three {\it ab initio} levels of theory MP2/aug-cc-pVTZ, CCSD(T)/cc-pVTZ and CCSD(T)-F12/cc-pVTZ-F12. It can be noted that, in general, the equilibrium rotational constants obtained from MP2/aug-cc-pVTZ and CCSD(T)/cc-pVTZ using Gaussian are closer for some species like A2, C1. The values obtained from CCSD(T)-F12/cc-pVTZ-F12 using Molpro was added for qualitative comparison even though computationally its convergence thresholds differ. However, the general agreement shows how reliable these methods are in computing geometric parameters for similar molecular species.

In Table~\ref{Dipole_rot_comparison} we also present the ground vibrational state rotational constants. These were computed using the equilibrium rotational constants from CCSD(T)-F12/cc-pVTZ-F12 and the vibrational contribution  $\Delta B^{\text{vib}}$ derived from the vibration-rotation interaction parameters determined from the MP2/AVTZ force field:

\begin{equation}\label{B0-abinitio}
B_0 = B_e \text{(CCSD(T)-F12/VTZ-F12)} + \Delta B^{\text{vib}} \text{(MP2/AVTZ)}  ~~,
\end{equation}

\noindent which is applied for the three rotational constants $A_0$, $B_0$ and $C_0$.
To assess these results, next we compare the theoretical rotational constants from Eq.~(\ref{B0-abinitio}) with the available experimental ground state rotational constants, i.e., those from aminoacetonitrile A1 and methylcyanamide C1.
The experimental ground state rotational constants of aminoacetonitrile  A1 are
$A_0$=30246.4909(9)~MHz, $B_0$=4761.0626(1)~MHz and $C_0$=4310.7486(1)~MHz~\cite{Jiang2023}. Our calculations provided the differences about 247~MHz, 26~MHz and 26~MHz, respectively. For methylcyanamide C1 (or MCA), the experimental ground state rotational constants fitted for the transitions with symmetry $a+$ are 
$A_0$=36090(156)~MHz, $B_0$=4977.13(7)~MHz and $C_0$=4505.13(7)~MHz~\cite{bak1980microwave}. 
Our calculations from Eq.~(\ref{B0-abinitio}) differ to be about 1109~MHz, 16~MHz and 20~MHz, respectively.
It should be noted that the experimental-calculated difference of $\Delta A$ is very large. Nevertheless, the comparison of the theoretical  with the experimental value of $A_0$ is less definitive because its experimental rotational constant has a substantial uncertainty whereas $B_0$ and $C_0$ are well determined. In addition, this large difference can also arise from the fact that the experimental rotational constants were estimated only from a fit of a few transitions with symmetry either $a+$ or $a-$~\citep{bak1980microwave}. MCA has two large amplitude motions which splits each rotational transition into four, labeled as  $a+$, $a-$, $e+$ and $e-$.  The values of $B_0$ and $C_0$ for both symmetries $a+$ and $a-$ are similar but the values of $A_0$ are rather different. According to this comparison, a new measurement of the experimental spectrum of MCA would be convenient to determine the constant $A_0$ with accuracy. 

Concerning the results of the total dipole moments given in Table~\ref{Dipole_rot_comparison}, the two methods (MP2 and CCSD(T) ) are in good accordance, with dipole moment predictions. For instance, A1 has 2.54 and 2.85~D at MP2/AVTZ and CCSD(T)/cc-pVTZ respectively while experimentally it was determined to be 2.64~D. The A2 isomer has a dipole moment of 0.00~D due to its molecular symmetry. In contrast, the MCA isomer exhibits a dipole moment of 5.04~D at the MP2 level, in close agreement with the CCSD(T). MP2 values are generally lower compared to CCSD(T), widely considered gold standard in quantum chemical calculations~\citep{Raghavachari1989}.
The dipole moments of these five isomers (A1, A3, C1, C2 and C3) are relatively large and, according to these values, they are potentially observable species either in the atmosphere or the interstellar medium. The isomer A2 does not have dipole moment. We should highlight that our results are comparable with others available in the literature \cite{McKissick2025}. For A1 species, the experimental dipole moment is 2.640(7)~D~\cite{Pickett1973}, akin to the calculations obtained for MP2/aug-cc-pVTZ, to be 2.54~D, and the coupled cluster methods CCSD(T)/cc-pVTZ and CCSD(T)-F12/cc-pVTZ-F12 to be 2.85~D and 2.86~D, respectively. A2 
 with 0.00 D agrees with previous work\cite{McKissick2025}. For A3 isomer, the dipole moment was recently computed via B3LYP/aug-cc-pVTZ to be 2.78~D~\citep{McKissick2025} while our calculations resulted in 2.83~D at MP2/aug-cc-pVTZ and 3.00~D at CCSD(T)/cc-pVTZ. The good agreement in these comparisons supports the reliability of our results. 

According to the accuracy provided by the level of theory MP2/aug-cc-pVTZ in Table~\ref{Dipole_rot_comparison}, we have also provided the rotational and centrifugal distortion (quartic and sextic)  constants of the S-reduced Watson Hamiltonian at MP2/aug-cc-pVTZ level of theory. In Table~\ref{rot-trans-die} we report their values of the isomers A1, A2 and A3 and in Table~\ref{rot-centr} their values of isomers C1, C2 and C3. On the one hand, a comparison with the available experimental parameters  is relevant to estimate the degree of accuracy of the calculations. For aminoacetonitrile (A1 or AAN) species, we obtained $A_{0}$=29965.759~MHz, $B_{0}$=4739.798~MHz, and $C_{0}$=4287.742~MHz and the difference with the experimental parameters~\cite{Jiang2023} are $\Delta A\approx281$~MHz,  $\Delta B\approx21$~MHz, and $\Delta C\approx23$~MHz. 
For species C1, the experimental-MP2 differences for the rotational constants are $\Delta A\approx1099$~MHz,  $\Delta B\approx14$~MHz, and $\Delta C\approx18$~MHz (see the experimental rotational constants fitted for the transitions with symmetry $a+$~\cite{bak1980microwave}). 
As for the results given by the Eq.~(\ref{B0-abinitio}), the experimental-calculated difference of $\Delta A$ is very large. 
In both cases, for A1 and C1, the differences of our MP2 calculations with respect to the experimental values of the rotational constants $B_0$ and $C_0$ are between 14 to 23~MHz, close to those obtained with Eq.~(\ref{B0-abinitio}), whereas the differences for $A_0$ are larger, as it happens in other works~\citep{Senent2021,InostrozaPino2023}.

On the other hand, comparison of our MP2 spectroscopic parameters with higher-level theoretical results \cite{McKissick2025} is also valuable. McKissick et al. computed the rotational constants as well as the quartic and sextic distortion constants for isomers A2 and A3 at the F12-TcCR+DZ QFF and F12-TcC QFF levels of theory. For isomer A2, the differences between our MP2 parameters and those reported by McKissick et al. are less than 0.5\% for the ground-state rotational constants and approximately 2\%, 5\%, and 0.2\% for the quartic distortion constants $D_J$, $D_{JK}$, and $D_K$, respectively. For A3 species, the differences of the ground state rotational constants are of 1.0\% for $A_0$, 0.1\% for $B_0$ and of 0.3\% for $C_0$, and about 3\%, 0.3\%, and 0.08\% for $D_J$, $D_{JK}$, and $D_K$.

Results from other isomers with higher electronic energies (A4-C11)  are provided in table S6 of the supplementary material. For those, we provided the rotational and centrifugal distortion (quartic and sextic) constants of the S-reduced Watson Hamiltonian, calculated at MP2/aug-cc-pVTZ level of theory.

\begin{table}[ht]
\caption{\justifying Comparison of the equilibrium rotational constants (in MHz) and total dipole moments $\mu$ (in Debye) of the isomers A1, A2, A3, C1, C2, and C3 among the three \textit{ab initio} methods MP2/aug-cc-pVTZ, CCSD(T)/cc-pVTZ and CCSD(T)-F12/cc-pVTZ-F12}
\label{Dipole_rot_comparison}
\begin{tabular}{llllll}
\hline
Para. & \begin{tabular}[c]{@{}l@{}}MP2/\\ AVTZ\end{tabular} & \begin{tabular}[c]{@{}l@{}}CCSD(T)\\ /cc-pVTZ\end{tabular} & \begin{tabular}[c]{@{}l@{}}CCSD(T)-F12\\ /cc-pVTZ-f12\end{tabular} & \begin{tabular}[c]{@{}l@{}}Ground \\ State\\ rotational \\ constants\end{tabular} & \begin{tabular}[c]{@{}l@{}}Previous\\ studies\end{tabular} \\ \hline
 & A1 &  &  &  &  \\ \hline
\(A_e\) & 30063.98 & 30109.31 & 30097.70 & 29999.50 & \begin{tabular}[c]{@{}l@{}}$A_0 = 30248.57(70)$ $^a$\\ 30246.49(9) $^d$\end{tabular} \\
\(B_e\) & 4764.54 & 4761.76 & 4760.24 & 4735.50 & \begin{tabular}[c]{@{}l@{}}$B_0 = 4760.93(60)$ $^a$\\ 4761.06(1)$^d$\end{tabular} \\
\(C_e\) & 4314.79 & 4313.45 & 4311.99 & 4284.94 & \begin{tabular}[c]{@{}l@{}}$C_0 = 4310.77(60)$ $^a$\\ 4310.75(1)$^d$\end{tabular} \\
\( \mu \) & 2.54 & 2.85 &  &  & 2.64$^a$ \\ \hline
 & A2 &  &  &  &  \\ \hline
\(A_e\) & 51213.62  & 51213.62 &  51352.20 & 50603.54 & $A_0=50971.20{}^b$ \\
\( B_e \) & 4686.86  &  4686.86 &  4669.43 & 4658.24 & $B_0=4674.90 {}^b$ \\
\( C_e \) & 4293.90 & 4293.90 & 4280.23  & 4265.92 & $C_0=4282.5{}^b$ \\
\( \mu \) & 0.0 & 0.00 &  &  & 0.00$^b$ \\ \hline
 & A3 &  &  &  &  \\ \hline
\(A_e\) & 48828.47 & 49369.58 & 48807.93 & 48261.72 & 48398.40$^b$ \\
\( B_e \) & 4682.14 & 4647.39 & 4680.73 & 4650.73 & 4677.10$^b$ \\
\( C_e \) & 4272.45 & 4247.60 & 4271.12 & 4242.75 & 4265.70$^b$ \\
\( \mu \) & 2.83 & 3.00 &  &  & 2.78$^b$ \\ \hline
 & C1 &  &  &  &  \\ \hline
\(A_e\) & 34943.27 & 34943.27 & 34932.44 & 34980.64 & $A_0$ =36090.00(156)$^c$ \\
\( B_e \) & 4997.71 & 4997.71 & 4995.31 & 4960.74 & $B_0$ =4977.13(7)$^c$ \\
\( C_e \) & 4522.44 & 4522.44 & 4520.30 & 4485.15 & $C_0$ = 4505.13(7)$^c$ \\
\( \mu \) & 5.04 & 5.00 &  &  &  \\ \hline
 & C2 &  &  &  &  \\ \hline
\(A_e\) & 54322.08 & 53951.25 & 54307.01 & 55229.14 &  \\
\( B_e \) & 4477.51 & 4403.89 & 4475.39 & 4441.88 &  \\
\( C_e \) & 4296.35 & 4228.78 & 4294.32 & 4237.00 &  \\
\( \mu \) & 2.37 & 3.40 &  &  &  \\ \hline
 & C3 &  &  &  &  \\ \hline
\(A_e\) & 50149.97 & 49861.33 & 50131.85 & 49594.88 &  \\
\( B_e \) & 5072.00 & 5061.12 & 5069.54 & 5028.51 &  \\
\( C_e \) & 4606.15 & 4594.74 & 4603.97 & 4569.51 &  \\
\( \mu \) & 3.09 & 3.58 &  &  &  \\ \hline
\end{tabular}
\justifying
$^a$ Experimental ground state rotational constants by Ref.~\cite{Pickett1973}. 

$^b$ Calculated rotational constants and dipole moments for A2 and A3 computed using the F12-TcCR+DZ and F12-TcC QFFs, respectively by Ref.~\cite{McKissick2025}.

$^c$ Experimental rotational constant of MCA (a+ symmetry) with assumed geometry from a related molecule with similar skeleton (\ce{CH3NHCl}).

$^d$ Experimental ground state rotational constants by Ref.~\cite{Jiang2023}
\end{table}

\begin{table}[]
\caption{\justifying Rotational and centrifugal distortion (quartic and sextic) constants of A1, A2, and A3 calculated at MP2/aug-cc-pVTZ level of theory}
\label{rot-trans-die}
\begin{tabular}{@{}lllllll@{}}
\toprule
\multirow{2}{*}{\begin{tabular}[c]{@{}l@{}}Mol.\\ Sym\end{tabular}} & \multicolumn{2}{c}{\begin{tabular}[c]{@{}c@{}}Rotational \\ Constants (MHz)\end{tabular}} & \multicolumn{4}{c}{\begin{tabular}[c]{@{}c@{}}Centrifugal Distortion\\ Constants\\ Watson S reductions\end{tabular}} \\ \cmidrule(l){2-7} 
 & \multicolumn{1}{c}{Equilibrium} & \multicolumn{1}{c}{g.s.} & \multicolumn{2}{c}{\begin{tabular}[c]{@{}c@{}}Quartic terms\\ (kHz)\end{tabular}} & \multicolumn{2}{c}{\begin{tabular}[c]{@{}c@{}}Sextic terms\\ (Hz)\end{tabular}} \\ \midrule
\multirow{7}{*}{A1} & $A_e$=30063.977 & $A_{0}$=29965.759 & $D_J$=3.1319 &  & $H_J$=0.0105 &  \\
 & $B_e$=4764.538 & $B_{0}$=4739.798 & \multicolumn{2}{l}{$D_{JK}$=-60.2600} & \multicolumn{2}{l}{$H_K$=-83.7900} \\
 & $C_e$=4314.793 & $C_{0}$=4287.742 & \multicolumn{2}{l}{$D_K$=723.1999} & \multicolumn{2}{l}{$H_{JK}$=-0.9714} \\
 &  &  & \multicolumn{2}{l}{$d_1$=-0.6935} & \multicolumn{2}{l}{$H_{KJ}$=14.7300} \\
 &  &  & \multicolumn{2}{l}{$d_2$=-0.02732} & \multicolumn{2}{l}{$h_1$=0.0036} \\
 &  &  &  &  & \multicolumn{2}{l}{$h_2$=-0.0003} \\
 &  &  &  &  & \multicolumn{2}{l}{$h_3$=0.0000} \\ \midrule
A2 & $A_e$=51213.618 & $A_{0}$=50626.262 & $D_J$=0.0009 &  & $H_J$=0.0001 &  \\
 & $B_e$=4686.857 & $B_{0}$=4659.622 & \multicolumn{2}{l}{$D_{JK}$=-0.0076} & \multicolumn{2}{l}{$H_K$=4.612} \\
 & $C_e$=4293.897 & $C_{0}$=4267.255 & \multicolumn{2}{l}{$D_K$=0.3581} & \multicolumn{2}{l}{$H_{JK}$=-0.0254} \\
 &  &  & \multicolumn{2}{l}{$d_1$=-0.0001} & \multicolumn{2}{l}{$H_{KJ}$=-0.1639} \\
 &  &  & \multicolumn{2}{l}{$d_2$=-0.4568D-05} & \multicolumn{2}{l}{$h_1$=0.0000} \\
 &  &  &  &  & \multicolumn{2}{l}{$h_2$=0.0000} \\
 &  &  &  &  & \multicolumn{2}{l}{$h_3$=$h_3$=0.0000} \\ \midrule
A3 & $A_e$=48828.471 & $A_{0}$=48282.260 & $D_J$=0.9608 &  & $H_J$=0.00009 &  \\
 & $B_e$=4682.136 & $B_{0}$=4652.139 & \multicolumn{2}{l}{$D_{JK}$=-6.8340} & \multicolumn{2}{l}{$H_K$=4.7610} \\
 & $C_e$=4272.453 & $C_{0}$=4244.087 & \multicolumn{2}{l}{$D_K$=312.8000} & \multicolumn{2}{l}{$H_{JK}$=-0.0149} \\
 &  &  & \multicolumn{2}{l}{$d_1$=-0.1089} & \multicolumn{2}{l}{$H_{KJ}$=-0.3547} \\
 &  &  & \multicolumn{2}{l}{$d_2$=-0.0054} & \multicolumn{2}{l}{$h_1$=0.0000} \\
 &  &  &  &  & \multicolumn{2}{l}{$h_2$=0.0000} \\
 &  &  &  &  & \multicolumn{2}{l}{$h_3$=0.0000} \\ \midrule
\end{tabular}
\end{table}

\begin{table}[ht]
\caption{\justifying Rotational and centrifugal distortion (quartic and sextic)  constants of C1, C2 and C3 calculated at MP2/aug-cc-pVTZ level of theory}
\label{rot-centr}
\begin{tabular}{@{}lllllll@{}}
\toprule
\multirow{2}{*}{\begin{tabular}[c]{@{}l@{}}Mol\\ sym\end{tabular}} & \multicolumn{2}{c}{\begin{tabular}[c]{@{}c@{}}Rotational \\ Constants (MHz)\end{tabular}} & \multicolumn{4}{c}{\begin{tabular}[c]{@{}c@{}}Centrifugal Distortion\\  Constants\\ Watson S reduction\end{tabular}} \\ \cmidrule(l){2-7} 
 & Equilibrium & \multicolumn{1}{c}{g.s.} & \multicolumn{2}{c}{\begin{tabular}[c]{@{}c@{}}Quartic terms\\  (kHz)\end{tabular}} & \multicolumn{2}{c}{\begin{tabular}[c]{@{}c@{}}Sextic terms\\  (Hz)\end{tabular}} \\ \midrule
C1 & $A_e$=34943.268 & $A_{0}$=34991.465 & \multicolumn{2}{l}{$D_J$=3.043} & \multicolumn{2}{l}{$H_J$=0.9714D-02} \\
 & $B_e$=4997.712 & $B_{0}$=4963.140 & \multicolumn{2}{l}{$D_{JK}$=-73.19} & \multicolumn{2}{l}{$H_K$=0.7418D+02} \\
 & $C_e$=4522.437 & $C_{0}$=4487.285 & \multicolumn{2}{l}{$D_K$=1219.000} & \multicolumn{2}{l}{$H_{JK}$=-0.49919} \\
 &  &  & \multicolumn{2}{l}{$d_1$=-0.6842} & \multicolumn{2}{l}{$H_{KJ}$=-4.514} \\
 &  &  & \multicolumn{2}{l}{$d_2$=-0.01828} & \multicolumn{2}{l}{$h_1$=0.3963D-02} \\
 &  &  &  &  & \multicolumn{2}{l}{$h_2$=0.3608D-03} \\
 &  &  &  &  & \multicolumn{2}{l}{$h_3$=0.1045D-03} \\ \midrule
C2 & $A_e$=54322.084 & $A_{0}$=55244.215 & \multicolumn{2}{l}{$D_J$=2.392} & \multicolumn{2}{l}{$H_J$=0.7589D-02} \\
 & $B_e$=4477.514 & $B_{0}$=4444.001 & \multicolumn{2}{l}{$D_{JK}$=-126.9} & \multicolumn{2}{l}{$H_K$=0.2436D+04} \\
 & $C_e$=4296.351 & $C_{0}$=4267.693 & \multicolumn{2}{l}{$D_K$=6947.000} & \multicolumn{2}{l}{$H_{JK}$=-1.572} \\
 &  &  & \multicolumn{2}{l}{$d_1$=-0.4241} & \multicolumn{2}{l}{$H_{KJ}$=-256.3} \\
 &  &  & \multicolumn{2}{l}{$d_2$=0.04162} & \multicolumn{2}{l}{$h_1$=0.2627D-02} \\
 &  &  &  &  & \multicolumn{2}{l}{$h_2$=-0.3920D+02} \\
 &  &  &  &  & \multicolumn{2}{l}{$h_3$=0.1084D+01} \\ \midrule
\multirow{6}{*}{C3} & $A_e$=50149.968 & $A_{0}$=49613.003 & \multicolumn{2}{l}{$D_J$=1.0530} & \multicolumn{2}{l}{$H_J$=-0.1031D-03} \\
 & $B_e$=5072.001 & $B_{0}$=5030.971 & \multicolumn{2}{l}{$D_{JK}$=-6.568} & \multicolumn{2}{l}{$H_K$=0.5992D+01} \\
 & $C_e$=4606.150 & $C_{0}$=4571.691 & \multicolumn{2}{l}{$D_K$=379.0000} & \multicolumn{2}{l}{$H_{JK}$=-0.0175} \\
 &  &  & \multicolumn{2}{l}{$d_1$=-0.1258} & \multicolumn{2}{l}{$H_{KJ}$= -0.4463} \\
 &  &  & \multicolumn{2}{l}{$d_2$=-0.1258} & \multicolumn{2}{l}{$h_1$=0.1563D-04} \\
 &  &  &  &  & \multicolumn{2}{l}{$h_2$=0.3651D-05} \\
 &  &  &  &  & \multicolumn{2}{l}{$h_3$=0.1039D-05} \\ \bottomrule
\end{tabular}
\end{table}

\subsection{Methylcyanamide (MCA or C1) molecule}

The isomer MCA (or C1 in fig. \ref{fig-isomers}) is the second one in energy and has the largest dipole moment among all the isomers of \ce{C2H4N2} presented in this work (see Fig.\ref{fig-isomers}).
.

\begin{figure}[h !]
\centering
    \includegraphics[scale=0.26
    ]{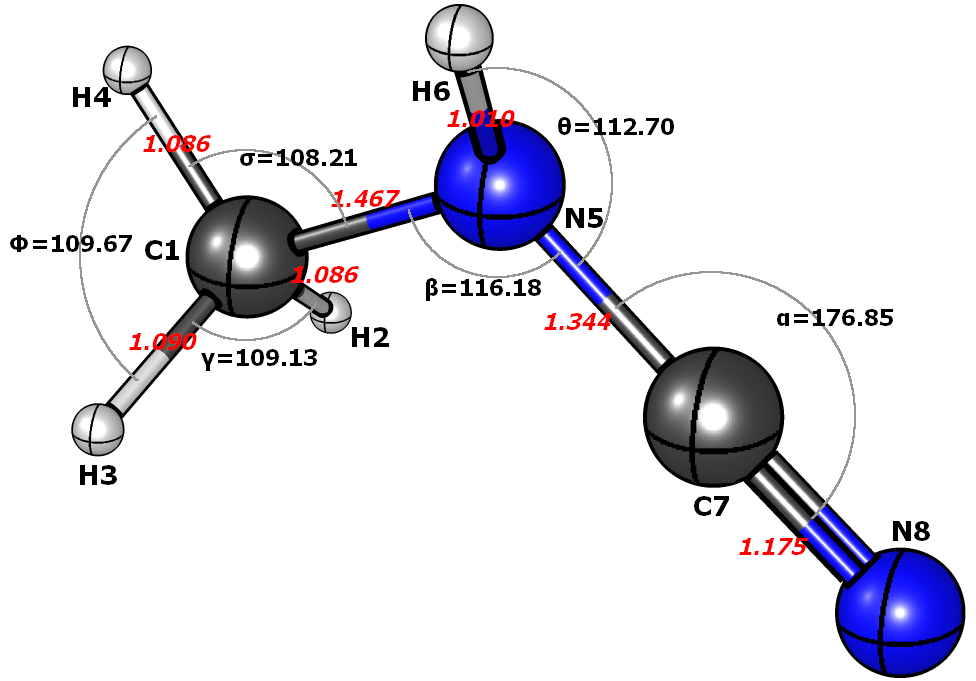}
    \caption{\justifying Optimised equilibrium geometry of \textit{MCA} (bond lengths (in \AA, red) and bond angles (in degrees, black)) calculated using CCSD(T)-F12/cc-pVTZ-F12}
    \label{cma-structure}
\end{figure}

MCA is a near prolate asymmetric top with an asymmetry parameter $\kappa = \frac{2 B_0 - A_0 - C_0}{A_0 - C_0}$=-0.969 computed with the ground state rotational constants reported in Table~\ref{Dipole_rot_comparison}.
Its equilibrium structure at the level of theory CCSD(T)-F12/cc-pVTZ-F12 is given in Table~\ref{tab-geom_xtion_aan_cma_etc-2}, and illustrated in Fig.~\ref{cma-structure}. The total dipole moment of MCA, computed at the level of theory CCSD(T)-F12/cc-pVTZ-F12, is the largest one (5.04~D) of the family \ce{C2H4N2} and its dipole components were computed to be $\mu_a=4.85$~D, $\mu_b=-1.08$~D, and $\mu_c= 0.89$~D, in good agreement with the two available experimental components $\mu_a$=4.72~D and $\mu_b$=1.30~D~\citep{bak1980microwave}.

MCA has 18 vibrational modes, of which at least two of them are large amplitude motions: the methyl internal rotation ($\nu_{18}$) and the inversion motion H6-N5 ($\nu_{16}$). The Table~\ref{tab-harmfreq-CCSDT-2} shows the harmonic and anharmonic contributions obtained at CCSD(T)/cc-PVTZ and MP2/AVTZ force field, respectively. The methyl group is an internal rotor with a frequency of 
95~cm$^{-1}$ and the inversion mode has a frequency of 420~cm$^{-1}$. These two large-amplitude modes split each rotational transition into four, hindering spectral characterization (see Ref. \cite{bak1980microwave}). An analysis of the torsional potential surface and its correlation with the N-H inversion angles is critical to determine the torsional and inversion barriers to quantify the splittings over the energy levels~\citep{ElHadki2022}. 

\begin{figure} 
\hspace*{-0.7cm}
\includegraphics[scale=0.50]{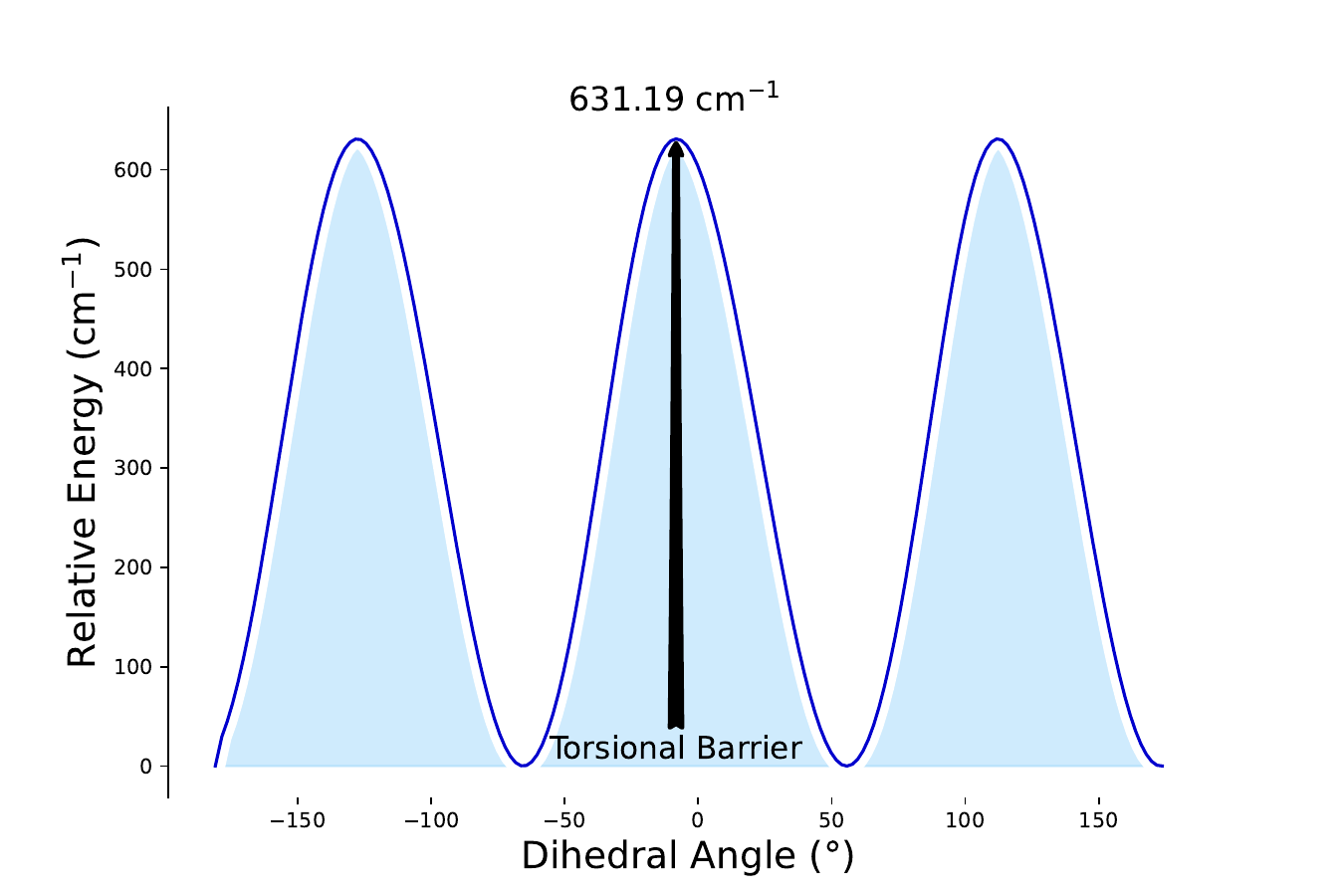}
    \caption{\justifying Torsional potential energy surface of \ce{CH3} top of the species MCA computed at the MP2/aug-cc-PVTZ level of theory with respect to the dihedral angle H4C1N5H6.} 
    \label{pes}
\end{figure}

In this work, the torsional potential surface of the large-amplitude motion of the methyl top of the C1 conformer has been calculated with respect to the dihedral angle H4C1N5H6 at the MP2/aug-cc-PVTZ level of theory. It has been computed by fixing the angle C7N5H6 at equilibrium, allowing for the free rotation of the methyl internal rotor. 
The resulting torsional potential, shown in Fig.~\ref{pes}, yields a barrier height of approximately 631~cm$^{-1}$. The only experimental estimate available for this parameter, 262~cm$^{-1}$ was obtained through an experimental procedure \cite{bak1980microwave}, and is significantly lower than the present theoretical value. This discrepancy is also evident when compared with the torsional barrier 1322~cm$^{-1}$ from \ce{CH3NHCl}, which share a similar molecular framework. The experimental value showed better agreement with molecular structures, such as \ce{CH3-N=N+=N-} with a barrier of 250~cm$^{-1}$.  Further experimental and theoretical investigations are therefore warranted to establish a more accurate determination of this barrier.

\subsection{Reaction Pathways}

\begin{figure}[htbp]
    \centering
    \includegraphics[width=0.8\linewidth]{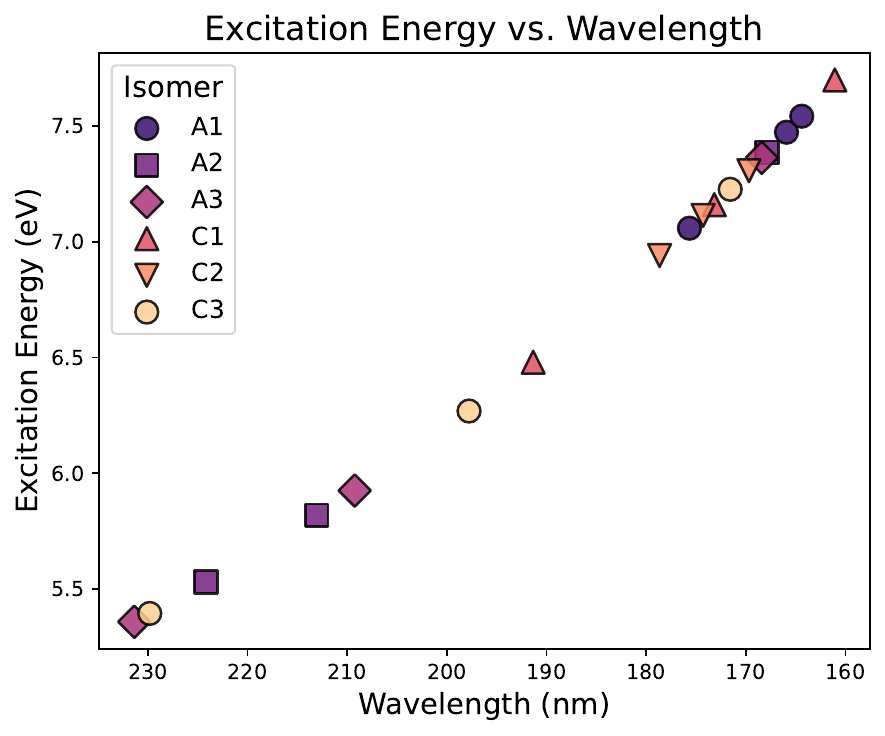}
    \caption{\justifying Illustration of trends in the electronic excitations between the first row isomers (A1-A3 and C1-C3)}
    \label{Excit_energies}
\end{figure}

In this section, before addressing the calculation of the reaction pathways, we compute the excitation energies, the linestrengths and the dipole moments of the isomers A1, A2, A3, C1, C2 and C3 using the  Single Excitation Configuration Interaction (CIS).
Figure \ref{Excit_energies} shows a simplified comparison of the trends of the electronic excitation energies, in eV and wavelength, for the isomers A1-A3, and C1-C3, providing insights on their absorption behaviour. Based on the table \ref{Excitation_energies}, it can be noted in Fig. \ref{Excit_energies} that the lowest excited electronic states, at the bottom-left, belong to A3, A2, C3 whereas the highest excited state, at the top-right, is from C1.
Table \ref{Excitation_energies} reports the calculated excitation energies, wavelengths and oscillator strengths of the first three excited states of A1, A2, A3, and C1, C2 and C3 isomers. It can be highlighted that isomers A2, C3 and A3 present the highest intensities of the transitions to the excited state, of which the third excited state of A2 has the highest intensity (0.8495), the third state of C3 has a linestrength of 0.7883 and the third state of A3 has an intensity of 0.2718. Table \ref{Excitation_energies} also shows the third electronic states of C3, A2 and A3 have the highest dipole moments with values of 2.9650~au, 2.1702~au and 1.2932~au, respectively. In particular, for the first three excited states of C1, C2 and C3 isomers, the energies of the first three excited states of C1 are computed as 6.48, 7.16, and 7.70~eV,  for C2 they are 6.94, 7.11, and 7.31~eV, and for C3 they are 5.39, 6.27, and 7.23~eV.  All these excited states absorb photons in the UV region, with wavelengths less than 230~nm, corresponding with the vacuum ultraviolet (VUV) absorption whose wavelengths $<200$~nm~\citep{Fleury2025}. These underscores possible photochemical dissociation forming reactive species that can recombine or contribute to complex chemical reaction networks via the absorption of vacuum ultraviolet (VUV) in the ISM  or in planetary atmospheres~\citep{Chang2023}. The intensities are weaker for the first two excited states of all isomers. However, the third electronic transition of A2, A3, 
 and C3 have relatively high linestrengths. In addition, as the first excited state of A3 is 5.3589 eV, C3 is at 5.3949~eV, while A2 is5.5304 eV, A3 has higher probability of absorbing photon at that energy with oscillator strenght of 0.0027. It is expected that A2, A3 and C3 are top candidate that are most likely UV-photochemically active species in the first and third excited states amongst the 6 isomers.

\begin{table}[ht]
\caption{Determination of excitation energies of the first row isomers (A1, A2, A3, C1, C2, and C3) of \ce{C2H4N2} computed using configuration interaction singles (CIS)}
\label{Excitation_energies}
\begin{tabular}{lllllll}
\hline
Isomer & \begin{tabular}[c]{@{}l@{}}Excited\\ State\end{tabular} & \begin{tabular}[c]{@{}l@{}}Elec.\\ state\end{tabular} & \begin{tabular}[c]{@{}l@{}}Excita. \\ Energy\\ (eV)\end{tabular} & \begin{tabular}[c]{@{}l@{}}Wave\\ length \\ (nm)\end{tabular} & \begin{tabular}[c]{@{}l@{}}Oscillator \\ Strength\end{tabular} & \begin{tabular}[c]{@{}l@{}}Dipole \\ Strength (au)\end{tabular} \\ \hline
A1 & 1 & $^1A''$ & 7.0583 & 175.66 & 0.0001 & 0.0003 \\
 & 2 & $^1A'$ & 7.4726 & 165.92 & 0.0106 & 0.0578 \\
 & 3 & $^1A''$ & 7.5419 & 164.39 & 0.0007 & 0.0035 \\
A2 & 1 & $^1A_u$ & 5.5304 & 224.19 & 0.0027 & 0.0200 \\
 & 2 & $^1B_g$ & 5.8190 & 213.07 & 0.0000 & 0.0000 \\
 & 3 & $^1B_u$ & 7.3850 & 167.89 & 0.8652 & 4.7819 \\
A3 & 1 & $^1A''$ & 5.3586 & 231.37 & 0.0007 & 0.0056 \\
 & 2 & $^1A''$ & 5.9253 & 209.25 & 0.0078 & 0.0539 \\
 & 3 & $^1A'$ & 7.3625 & 168.40 & 0.8495 & 4.7096 \\
\begin{tabular}[c]{@{}l@{}}C1\\ (MCA)\end{tabular} & 1 & $^1A$ & 6.4797 & 191.34 & 0.0004 & 0.0023 \\
 & 2 & $^1A$ & 7.1598 & 173.17 & 0.0036 & 0.0206 \\
 & 3 & $^1A$ & 7.6986 & 161.05 & 0.0021 & 0.0113 \\
C2 & 1 & $^1A$ & 6.9398 & 178.66 & 0.0003 & 0.0020 \\
 & 2 & $^1A$ & 7.1135 & 174.29 & 0.0003 & 0.0739 \\
 & 3 & $^1A$ & 7.3065 & 169.69 & 0.0092 & 0.0511 \\
C3 & 1 & $^1A''$ & 5.3949 & 229.82 & 0.0003 & 0.0023 \\
 & 2 & $^1A''$ & 6.2688 & 197.78 & 0.0029 & 0.0188 \\
 & 3 & $^1A'$ & 7.2263 & 171.57 & 0.7883 & 2.9650 \\ \hline
\end{tabular}
\end{table}

Next, we studied the formation and dissociation pathways using the intrinsic reaction coordinate (IRC). We have ensured that every transition state is a first-order saddle point, whose reaction coordinate $\rho$ is characterised by
\begin{equation}\label{ts-eq}
    \frac{\partial^2 W}{\partial \rho^2} 
    \begin{cases} 
      < 0, & \text{for transition states} \\
      > 0, & \text{for all other points (initial and final)}
    \end{cases}
\end{equation}

\noindent where W is the potential energy~\citep{Fukui1981}.
The forward reactions were characterised by a positive transition vector components whereas the backward reactions were characterised by negative transition vector components in the direction of the reaction coordinate.

\begin{table*}[]
\caption{Formation and destruction pathways of \ce{CH3NHCN} provided with their activation energies  and enthalpies (in kJ$\sim$mol$^{-1}$ and eV) computed at B3LYP/6-31g(d,p)//CCSD(T)/aug-cc-pVTZ levels}
\label{Mech-table}
\begin{tabular}{lllccc}
\hline
\multicolumn{1}{c}{\multirow{2}{*}{\begin{tabular}[c]{@{}c@{}}Reaction \\ Number\end{tabular}}} & \multicolumn{1}{c}{\multirow{2}{*}{Reaction}} & \multicolumn{1}{c}{\multirow{2}{*}{Type}} & \multirow{2}{*}{\begin{tabular}[c]{@{}c@{}}Activation energy\\ kJ mol$^{-1}$ (eV)\end{tabular}} & \multirow{2}{*}{\begin{tabular}[c]{@{}c@{}}Enthalpy of reaction\\ kJ mol$^{-1}$ (eV)\end{tabular}} & \multirow{2}{*}{\begin{tabular}[c]{@{}c@{}}Overall \\ reaction\end{tabular}} \\
\multicolumn{1}{c}{} & \multicolumn{1}{c}{} & \multicolumn{1}{c}{} &  &  &  \\ \hline
R1 & CHNHCN + H$_2$ $\rightarrow$ CH$_3$NHCN & Form. & 49.40 (0.51) & -365.57 (-3.78) & Exo \\
R2 & CH$_2$NCN + H$_2$ $\rightarrow$ CH$_3$NHCN & Form. & 205.50 (2.13) & -300 (-3.11) & Exo. \\
R3 & CH$_3$NHCN $\rightarrow$ CH$_2$NH + HCN & Dest. & 430.12 (4.45) & 84.47 (0.87) & Endo. \\
R4 & CH$_3$NHCN $\rightarrow$ CH$_2$NH$_2$CN & Dest. & 387.87 (4.02) & -341.83 (-3.54) & Exo. \\
R5 & (CH$_3$)$_2$NCN $\rightarrow$ CH$_3$NHCN + CH$_2$ & Form. & 350.24 (3.63) & 423.75 (4.39) & Endo. \\
R6 & CH$_3$NH + CN $\rightarrow$ CH$_3$NHCN & Form. & 0.0 &  & Barrierless \\
R7 & CH$_3$NHCN + H$_2$ $\rightarrow$ CH$_3$NH$_2$ +HCN & Dest. & 335.76 (3.48) & -28.61 (-0.29) & Exo. \\
R8 & CH$_3$NHCN $\rightarrow$ CH$_3$NCNH & Dest. & 347.34 (3.60) & 80.32 (0.83) & Endo. \\
R9 & CH$_3$NHCN + H $\rightarrow$ CH$_3$NHCNH & Dest. & 31.84 (0.33) & -100.90(-1.04) & Exo. \\
R10 & OH + CH$_3$NH$_2$CN $\rightarrow$ CH$_3$NHCN + H$_2$O & Form. & 0.0 (0.0) &  & Barrierless \\
R11 & OH + CH$_3$NHCNH $\rightarrow$ CH$_3$NHCN + H$_2$O & Form. & 260.47 (2.69) & -44.64 (-0.462) & Exo. \\
R12 & CH$_3$NCN + H $\rightarrow$ CH$_3$NHCN & Form. & 0.0 &  & Barrierless \\
R13 & CH$_3$ + HNCN $\rightarrow$ CH$_3$NHCN & Form. & 0.0 &  & Barrierless \\
R14 & CH$_2$NH + HCN $\rightarrow$ CH$_3$NHCN & Form. & 386.04 (4.00) & -84.47 (-0.87) & Exo. \\ \hline
R15 & H$_2$ + HCNCNH $\rightarrow$ CH$_3$NCNH & Form. & 45.63 (0.47) & -232.03 (-2.40) & Exo. \\
R16 & CH$_2$NCHNH $\rightarrow$ CH$_3$NCNH & Form. & 322.97 (3.34) & -16.72 (0.17) & Exo. \\
R17 & CH$_3$NCN + H $\rightarrow$ CH$_3$NCNH & Form. & 0.0 &  & Barrierless \\ \hline
R18 & CH$_2$NCHNH $\rightarrow$ HNC + CH$_2$NH & Dest. & 320.02 (3.31) & 93.66 (0.97) & Endo. \\
R19 & CH$_2$NCN + H$_2$ $\rightarrow$ CH$_2$NCHNH & Form. & 319.75 (3.31) & -103.46 (-1.07) & Exo. \\
R20 & HCN + HNC + H$_2$ $\rightarrow$ CH$_2$NCHNH & Form. & 153.20 (1.58) & -192.26 (-1.99) & Exo. \\ \hline
R21 & cis-NHCHCHNH $\rightarrow$ CHNH$_2^+$ + HCN & Dest. & 208.61 (2.16) & 200.41 (2.38) & Endo. \\
R22 & NH2CCHNH $\rightarrow$ trans-NHCHCHNH & Form. & 219.68 (2.27) & -155.10 (-1.61) & Exo \\ \hline
\end{tabular}

\justifying Based on benchmarking of the chosen \textit{ab initio methods, an uncertainty of 3.0–4.5 kJ mol$^{-1}$ at the CCSD(T)/aug-cc-pVTZ level of theory \cite{West2023} can be expected}. Barrier heights for formation and dissociation of \ce{CH3NHCN} (C1) from R1 to R14; \ce{CH3NCNH} (C2), from R15 to R17; and \ce{CH2NCHNH} (C3), from R18 to R20. 1,2-di-iminoethane or NHCHCHNH represented by R21 and R22, respectively.
\end{table*}

\begin{figure*}{}
    \includegraphics[width=1\linewidth]{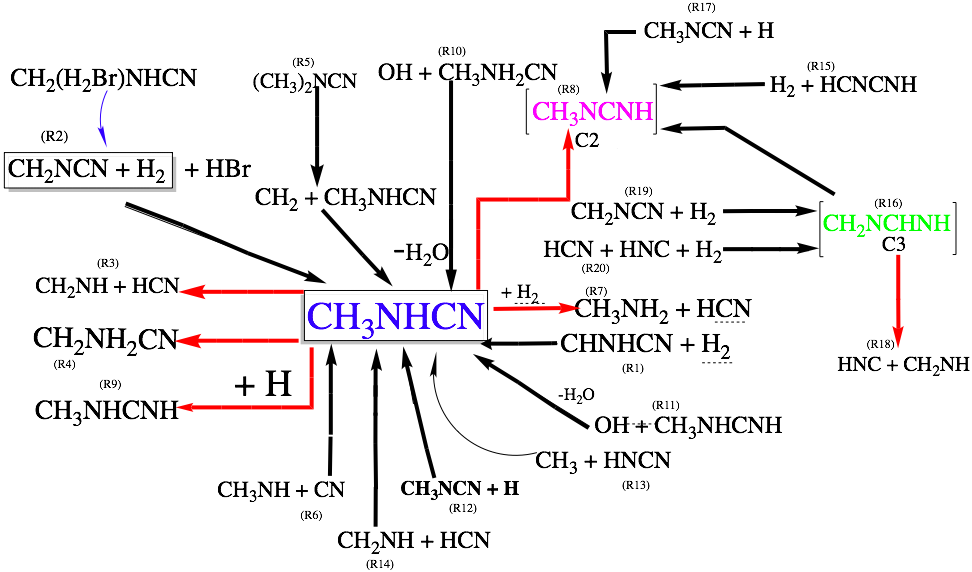}
    \caption{\justifying Gas-phase reaction network for the formation and destruction of MCA, C2 and C3. Each reaction is numbered, formation pathways are depicted by black arrows, destruction and pathways leading to other products are depicted by red arrows whereas intermediate path is shown with a blue arrow.}
    \label{reaction_network}
\end{figure*}

\begin{figure*}[]
    \centering
    \includegraphics[width=1.0\linewidth]{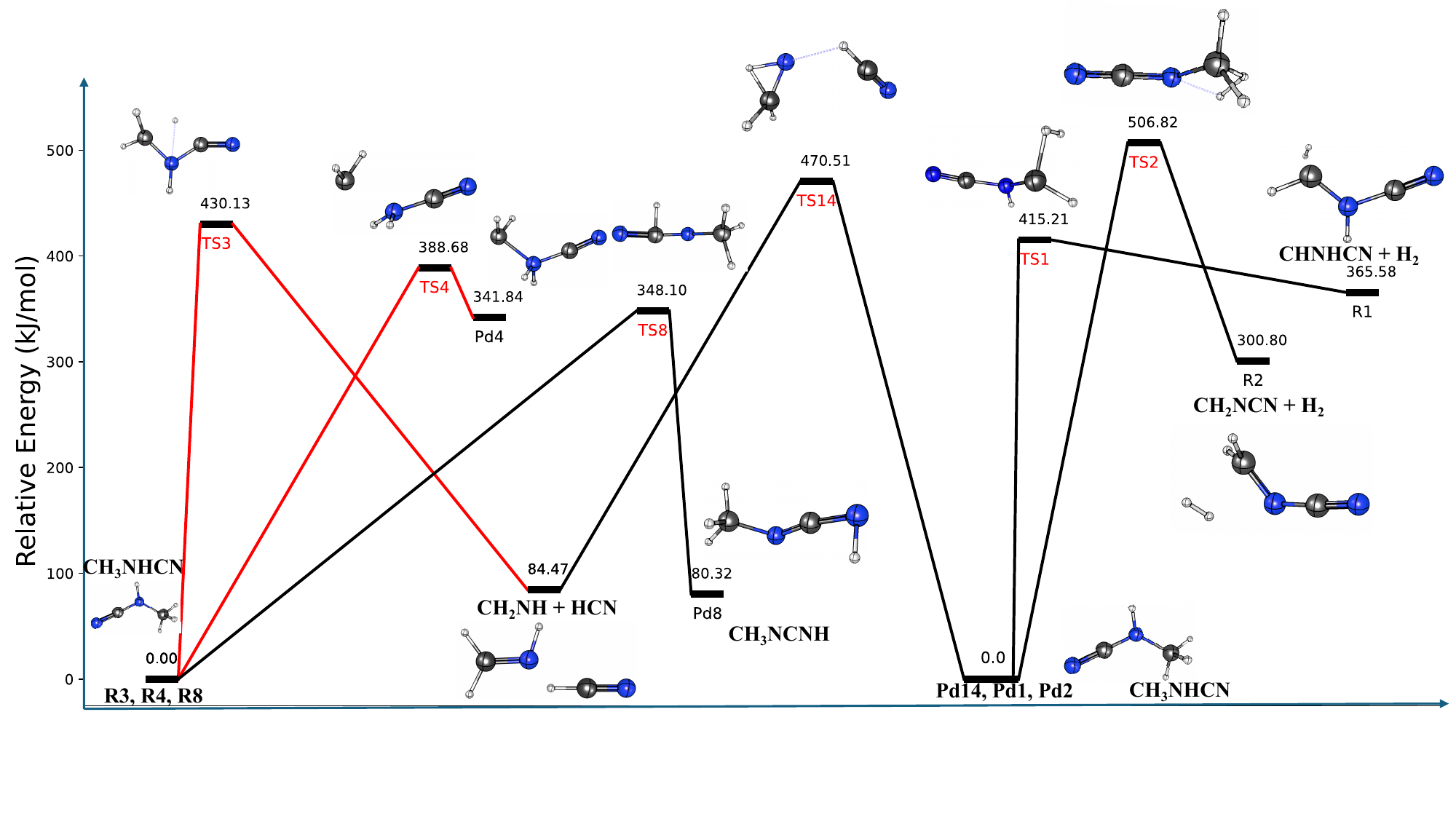}
    \caption{\justifying Some studied reaction pathways of formation and destruction of MCA computed at the B3LYP/6-31g(d,p)//CCSD(T)/aug-cc-pVTZ, where black lines corresponds to formation pathways and red lines corresponds to destruction pathways.}
    \label{rxn-mech}
\end{figure*}

Table \ref{Mech-table} shows the reaction paths leading to the formation and destruction of MCA and some of the other isomers of the \ce{C2H4N2} including their activation energies and enthalpies. Table \ref{Mech-table} presents the possible reactions involving the species MCA(C1), C2, and C3, as well as A2 and A3, and are organized according to them.
Most of the studied reaction pathways are illustrated in Figs. \ref{reaction_network} and \ref{rxn-mech}. Out of these reactions we constructed Fig. \ref{rxn-mech} with reference to \ce{CH3NHCN}. In Figure \ref{rxn-mech}, $R_n$ refers to the reaction number, $T_n$ designates the corresponding transition state, and the products of reaction are denoted by $Pd_n$ for the associated reaction. The reactions follow a pathway from reactant(s) (minima) through transition states (TSn, where n is the reaction number) to products.  The activation energy ($E_a$) for all the reactions studied here were computed.

In reaction R1—\ce{CH3NHCN} is formed via the neutral-neutral addition of molecular hydrogen (\ce{H2}) to CHNHCN. The transition state (TS), with an activation energy of \(49.40~\text{kJ mol}^{-1}\), is characterised by an elongated N–H bond and an emerging H–H distance. This indicates a concerted hydrogen abstraction in which one hydrogen atom, initially bonded to a nitrogen center, migrates toward a second hydrogen atom nearby, forming a transient dihydrogen (\ce{H2}) moiety. This is a significant process, as neutral-neutral reactions contribute to the formation of COMs, especially in the ISM \cite{Mendoza2025}.

Reaction R2, there was an  electrostatic attraction between molecular hydrogen (\ce{H2}) and Nitrogen of \ce{CH2NCN}  and the other hydrogen atom with carbon of -\ce{CH2}-moeity of \ce{CH2NCN}. The process led to the formation of a transition state with a high activation energy (205.50~{kJmol$^{-1}$)} then followed by elongation of the hydrogen molecule and bringing about a homolytic cleavage, hydrogen transfer then lead to stabilisation and formation of \ce{CH3NHCN}.
Similar or different processes can be seen for other reactions such as R3; atoms rearrangement followed by proton transfer and subsequent N-C bond cleavage to yield the dissociation products etc. Some of these transition states through each reaction is shown in figure \ref{rxn-mech}. The plot showcases formation/destruction pathways of C1 that are related by species that have been detected in the ISM, for example, \ce{CH2NCN} in R2 pathway (\ce{CH2NCN + H2 \rightarrow CH3NHCN}) has been detected in the Galactic Center molecular cloud G+0.693-0.027 \cite{San_Andres2024-st}. The reactant in R1 corresponding to \ce{CHNHCN} is an isomer of \ce{CH2NCN} and \ce{H2} is and abundant specie in the universe and detected in the ISM \cite{Wakelam2017}, and R3 and R14 are connected by dissociation product HCN + \ce{CH2NH}, both are detected species in different regions such as Sagittarius B2 (Sgr B2) \cite{Gorski2021,Neumann2024}.  It also allows the comparison of the relative energies of other reactants and transition states with respect to \ce{CH3NHCN}, some of these reactions related to the \ce{CH3NHCN} are depicted. The red lines illustrate destruction pathway while the blue lines correspond to the formation pathways.

It is noted that there are some barrierless reactions (the reactions involves radical whose unpaired electrons makes them readily reactive without the need to overcome activation energy) leading to C1 isomer as a product  such as the reactions R6, R10, R12 and R13. The reaction  with the highest activation energy corresponds to reaction R3 and R14 with R3 being  429.35 kJ~mol$^{-1}$ at coupled-Cluster level of theory while R14 has 386.04 kJ~mol$^{-1}$ at CCSD(T)/aug-cc-pVTZ. \ce{CH2NH} and HCN are species commonly encountered during the formation and dissociation of \ce{CH3NHCN} as indicated by the forward and reverse reactions R3 and R14 passes through different transition states as illustrated in Figure \ref{rxn-mech}, while R3 is an endothermic dissociation pathway , R14 is a exothermic formation pathway. However, the large activation energy in both reactions is not typical in the ISM. Therefore, such reactions  would require energetic process (e.g., UV irradiation, cosmic rays) or catalytic surfaces. 
In terms of the enthalpy of the reactions, exothermic reactions are R1, R2, R4, R7, R9, R11, R14 for C1; R15 and R16 for C2; R19 and R20 for C3; and R22 for A2. It is worthy to note that most of these exothermic reactions are formation pathways whereas the endothermic reactions involving C1 species are those numbered as R3, R5, and R8;  R18 involving C3; and R21 for A3. Reaction R5 is a decomposition  reaction of cyanodimethylamine yielding MCA and methylene. This reaction, which occurs via a hydrogen shift to the amine nitrogen followed by cleavage of the $-$CH$_2$ group, is characterised by an endothermic enthalpy CCSD(T)/aug-cc-pVTZ energy barrier of 423.75 kJ~mol$^{-1}$ (1.74 eV)).  
According to the spin selection rules for allowed and forbidden reactions \citep{Harvey2007-hc}, which forbid transitions between states with different total spin and different spin multiplicity, all reactants involved in the formation of MCA (first section of Table~\ref{Mech-table}) occur in the singlet state ($^1\Sigma$), similarly to the product (MCA). In summary, it can be inferred from our analysis that several reaction generating MCA, as well as species C2, C3, A2 and A3, are potentially feasible. 

\subsection{Atmospheric and Astrophysical implications}

The detection of several key reactants —such as methanimine (CH$_2$NH), hydrogen cyanide (HCN), and the hydroxyl radical (OH) \citep{EPSC2015-4,Bruno2023,Heard2006}—  in various layers of Earth’s atmosphere supports the current hypotheses regarding the atmospheric roles of the molecular species studied in Table \ref{Mech-table}. For example, the
 dissociation products in reaction R3 are methanimine (\ce{CH2NH}) and hydrogen cyanide (HCN). Methylene imine (\ce{CH2NH}) has been found in the ionospheric layer of the Earth's atmosphere \citep{EPSC2015-4} whereas HCN is found in both the troposphere and stratosphere of the Earth \citep{Cicerone1983}. This implies that, if present in the atmosphere, MCA would undergo decomposition to methylene imine and HCN. HCN is a known toxic compound found in different parts of the atmosphere \cite{Cicerone1983}.  Reaction R5 shows MCA can also be obtained from the decomposition of dimethylcyanamide \ce{(CH3)_2NCN} into MCA and methylene \ce{CH2}, which is a highly reactive intermediate and a part of many compounds such as drugs, fossil fuels and plastics via its ability to act as an adduct \citep{Shavitt1985-ig}. \ce{CH2} is present in methane-oxygen and ethylene-oxygen flames \citep{Peeters1975}. 
 Reactions R1 and R9, with low activation energies relatively close to the thermal energy available in the atmosphere (\(\sim 3.7 \, \text{kJ/mol}\)), could be expected feasible in the lower layers of the atmosphere.
The exothermic reaction R7, from MCA, gives  methylamine (\ce{CH3NH2}) as product, which is detected in the atmosphere \citep{Yu2014-yk} and could be present in the upper atmosphere of Titan, reacting with CN radicals to form MCA and other N-bearing organic compounds~\citep{Sleiman2018}.
 In addition, there are four barrierless reactions R6, R10, R12 and R13 forming the conformer C1 as well as another formation reaction R17 of confomer C2.

In astrophysical environments, N-bearing molecules are among the most abundant species in sources such as the Taurus Molecular Cloud (TMC-1) \citep{Chen2022}. The new finding of N-bearing molecules in star-forming regions,  protoplanetary disks, and other interstellar sources is relevant for paving the way for the astrochemical understanding and modeling of the reaction mechanisms in dust-grains and in the gas phase processes \cite{Mondal2023}. 
It is expected that, apart from the barrierless reactions, the other reactions studied in Table~\ref{Mech-table} 
could also occur in the ISM sources. For instance, the reaction R1 coming from the association between CHNHCN and hydrogen radicals has a low activation energy of 0.51 eV. This kind of reactions is relevant in the ISM especially in ice-surfaces or dust grains \cite{Wakelam2017,Cui2024}.  In reaction R2, two hydrogen radicals add up to the \ce{CH2} and the middle N atom of N-cyanomethanimine (\ce{CH2NCN}) forming \ce{CH3NHCN}. The compound \ce{CH2NCN} has been already detected in space \citep{San_Andres2024-st} and its isomer CHNHCN in reaction R1 can be formed by tautomeric transfer of a hydrogen from the carbon atom to the nitrogen atom.
  MCA (\ce{CH3NHCN}) dissociates into (\ce{CH2NH}) and the HCN in reaction R3, which products have been identified in ISM sources~\cite{Dickens1997,Gorski2021,Mendoza2025} as well as \ce{CH2NH} is known to play a vital role in prebiotic chemistry and the formation of amino acid  \cite{Zhu2019}. Reaction R4 is basically an isomerisation reaction between \ce{CH3NHCN} to \ce{CH2NH2CN}, involving a transfer of one hydrogen atom from the methyl group (\ce{CH3}) to \ce{NH} group. In addition, the reactions  with high barrier heights in Table~\ref{Mech-table}
  could also occur in moderately high kinetic temperatures environments, such as Galactic center molecular clouds, or in regions with higher energy processes, radiation and/or shock fronts, provided that there is an additional energy source, like UV radiation or catalyst, i.e., the reaction could be triggered at any environment when there are appropriate chemical and physical conditions \citep{Inostroza-Pino2024-qb}.
  
The present results provide new pathways and thermochemical data for N-bearing reactions for which no previous formation pathways or barrier information was available. These reactions, particularly those involving \ce{CH2NH}, \ce{CH3NH}, and \ce{HCN} derivatives, are relevant to the formation of methylcyanamide (MCA) and other related species, which have been proposed as precursors of interstellar glycine. The calculated reaction enthalpies and activation energies can be incorporated into astrochemical reaction networks such as UMIST \citep{Millar2023} and KIDA \citep{Wakelam2024} to improve the modeling of nitrogen chemistry in hotcores, dense molecular clouds, and protoplanetary disks. Inclusion of these pathways/thermochemical parameters will refine gas-gas or gas-grain code for abundance estimation \citep{Iqbal2018,Ge2021-yu} enhance the predictive capacity of interstellar chemical models. MCA (\ce{CH3NHCN}) and other conformers A1, A2, A3, C2 and C3 have absorption wavelengths within the range critical for photochemistry (115–230~nm)  (see Tab.~\ref{Excitation_energies}). Thus, they can be involved in photochemical reactions of interest to determine the composition of gas giant exoplanetary atmospheres~\citep{Fleury2025}. The James Webb Space Telescope (JWST) observed the evidence of photochemistry and its impact in the atmospheres of exoplanets like the gas giant WASP-39b~\citep{Tsai2023}. 

Besides, the thermal dependence of VUV absorption cross sections of molecules are not well known at the characteristic temperatures of the observed exoplanet atmospheres \cite{Fleury2025} and, therefore, this study can also hint at the photochemical processes of these molecules on them. Most of the reactions in Table~\ref{Mech-table} have high activation energies but below the excitation energies of the conformers A1-A3 and C1-C3 provided in Table~\ref{Excitation_energies}. Consequently, it could be expected that a combination of the photochemistry processes and the reactions studied here could produce the conformers of \ce{C2H4N2} and other molecular species key in the exoplanetary atmospheres.

\section{Conclusions} \label{concl}

This comprehensive study of the family of isomers  \ce{C2H4N2} includes the {\it ab initio} calculation of their electronic ground state and excitation energies, structural geometries, spectroscopic rotational and vibrational parameters, dipole moments, and the activation energies and enthalpies in some reaction pathways involving conformers C1, C2 and C3. The electronic energies for all the conformers have been computed using the high level CCSD(T)-F12/cc-pVTZ-F12 and spectroscopic parameters using correlated \textit{ab initio} methods CCSD(T)/cc-pVTZ and MP2/aug-cc-pVTZ. These have shown a good agreement with the available experimental and computed data. The dipole moments of the isomers A1-A4 and C1-C4 have been computed with significant values larger than $2$~D. In particular, the conformer methylcyanamide (C1 or MCA) has a notable large dipole moment of 5.04~D in comparison to other isomers of the family and is the next most energetically stable species lying at 0.27~eV (2177.70 $\text{cm}^{-1}$). From these two facts, we would expect plausible its radioastronomical detection.

 Excitation energies of the first three isomers of Cn show that C3 has the lowest value of 5.3949 eV compared to 6.4797 eV for MCA and, therefore, it is more reactive than MCA. 
 A study of the chemical network of MCA, C2 and C3 has been carried out to understand the formation and destruction pathways in the gas phase by computing the activation energies and the enthalpies at the level of theory B3LYP/6-31g(d,p)//CCSD(T)/aug-cc-pVTZ. The most probable pathways involve the barrierless processes and other neutral-neutral reactions with relatively low activation energies. 
The current investigation intends to provide the relevant information for the spectral characterization of the family of isomers \ce{C2H4N2}, of which spectra are incomplete or have never been measured yet. In addition, the calculation of the relevant quantities for the chemical pathway networks of this set of species will shed light on the abundance evolution patterns in environments such as the Earth's atmosphere and ISM after incorporating them into the atmospheric or astrochemical models. 
\section{Supplementary Material}
The supplementary material provides information on the remaining isomers that are not included in the main text. It includes three sections: The first section describes the relevant structural parameters of isomers with higher electronic energy; The second section presents the harmonic and anharmonic vibrational frequencies of isomers with higher electronic energy; and the third section reports the rotational and centrifugal distortion constants of the isomers.

\begin{acknowledgments}
      This work was supported by a PhD scholarship from the DCA, Universidad Autónoma de Chile. NI acknowledges FONDECYT grant Nº1241193 and VRIP.
      This project has also received funding from the European Union's Horizon 2020 research and innovation program under Marie Sklodowska-Curie grant agreement No. 872081, and grant PID2022-136228NB-C21 (M.C.) funded by MCIN/AEI/10.13039/501100011033, and, as appropriate, by "ERDF A way of making Europe", the "European Union", or the "European Union NextGenerationEU/PRTR". This work is also supported by the Consejería de Transformación Económica, Industria, Conocimiento y Universidades, Junta de Andalucía and European Regional Development Fund (ERDF 2021-2027) under the project EPIT1462023 (M.C. and E.M.).
      
\end{acknowledgments}

\bibliographystyle{aipnum4-1}
\bibliography{reference}

\end{document}